\documentclass{rQUF2e}

\usepackage{epstopdf}% To incorporate .eps illustrations using PDFLaTeX, etc.
\usepackage{subfigure}% Support for small, `sub' figures and tables
\usepackage{color}

\theoremstyle{plain}
\newtheorem{theorem}{Theorem}[section]

\newtheorem{proposition}[theorem]{Proposition}

\theoremstyle{definition}
\newtheorem{definition}{Definition}

\theoremstyle{remark}
\newtheorem{remark}{Remark}

\begin{document}

\title{  Is the variance swap rate affine in the spot variance? Evidence from S\&P500 data    }

\author{M.E. MANCINO $\dag$ , S. SCOTTI ${\ddag}$ and G. TOSCANO$^{\ast}$ $\S $\thanks{$^\ast$Corresponding author.
Email: giacomo.toscano@sns.it}\\
\affil{$\dag$Department of Economics and Management, University of Florence, Via delle Pandette 32, 50127 Florence, Italy\\
$\ddag$LPSM, Université de Paris,  rue Thomas Mann 5, 75205 Paris cedex 13, France \\
$\S$Scuola Normale Superiore, Piazza dei Cavalieri 7,  56126  Pisa, Italy} \received{submitted April 5, 2020} }

\maketitle

\begin{abstract}
We empirically investigate the functional link between the variance swap rate and the spot variance. Using S\&P500 data over the period 2006-2018, we find overwhelming empirical evidence supporting the affine link analytically found by  \cite{KMKV2011} in the context of exponentially affine stochastic volatility models. Tests on yearly subsamples suggest that exponentially mean-reverting variance models provide a good fit during  periods of extreme volatility, while polynomial models, introduced in  \cite{Cuchiero-PhD}, are suited for years characterized by more frequent price jumps.
\end{abstract}


\begin{classcode}C2, C12, C51, G12, G13.\end{classcode}

\section{Introduction}

The class of the exponentially affine processes, introduced in the seminal paper by   \cite{DufPanSin} and characterized by  \cite{F01}, has received large consensus in the quantitative finance literature, based on  its main advantages in terms of analytical tractability and empirical flexibility. The classic example of an exponentially affine process, and the only one with continuous paths, is the CIR
diffusion, see  \cite{CIR85}. The related stochastic volatility model, studied by   \cite{Heston}, is considered as a reference model by scholars and practitioners. 
  \cite{KMKV2011} have studied the valuation of options written on the quadratic variation of the asset price within the exponentially affine stochastic volatility framework. In particular, they have proved, analytically, the existence of an affine link between the expected cumulated variance, i.e., the variance swap rate, and the spot variance. Note that the class of stochastic volatility models considered in \cite{KMKV2011} allows for jumps and leverage effects, but fails to include some popular stochastic volatility models, e.g., the models by \cite{beckers1980constant,G2016, hagan2002managing,32}. 
The variance swap is possibly the most plain vanilla contingent claim written on the realized variance. Indeed, it can be seen, to some extent, as the forward of the integrated variance of log-returns (see, for instance, \cite{BSS19, CS07, CW08, FGM16, JMSZ19, KMKV2011}). Volatility derivatives
appear nowadays with a vast demand, especially after the global financial crisis of 2008, which induced large fluctuations in the volatility and other indicators of market stress. The large demand for volatility derivatives has resulted in a major increase in their liquidity, and thus in the reliability of their prices (see, for instance, \cite{CW08}).

Based on the study by   \cite{KMKV2011}, two natural questions arise: (i) could we analytically identify a wider class of models which admits an affine link between the variance swap rates and the spot variance? (ii) is it possible to test if empirical data satisfy a given link (e.g., affine, quadratic) between the variance swap rate and the unobservable spot variance?
This paper contributes to answering both questions. With regard to question (i), we prove that a larger class of models exhibits a linear link between the variance swap rate and the spot variance, and we show that a quadratic (respectively, affine) link appears between the variance swap rate and the multidimensional stochastic process characterizing the model, in the presence (respectively, absence) of jumps, within the class of polynomial models (see \cite{Cuchiero-PhD} and \cite{CKRT12}). With regard to question (ii), we set up a simple testing procedure, based on Ordinary-Least-Squares (OLS), in which the unobservable spot variance is replaced with efficient Fourier estimates thereof. Then, we apply it to S\&P500 empirical data over the period 2006-2018.

In particular, our first result is showing that a model exhibits the affine  link between the variance swap rate and the spot variance if the stochastic
differential equation satisfied by the latter is the sum of an affine drift and a zero-mean stochastic process. We term this class exponentially mean-reverting variance models.
This class is fairly large. In fact, it contains not only exponentially affine processes with jumps (see, e.g., \cite{Bates96, bnsmod, bns, DufFilSch, JMS17, JMSZ19}), but also, under suitable conditions (see \cite{CKRT12, AFP18}), polynomial processes. Moreover, it also contains some models based on the fractional Brownian motion, like the rough Heston model (see, for instance \cite{bayer2016pricing,ElEuch, el2019characteristic,gatheral2018volatility}).
However, it is worth noting that many popular models, e.g., the CEV model (\cite{beckers1980constant}), the SABR model (\cite{hagan2002managing}), the $3/2$ and $4/2$ models (\cite{32, G2016}), fail to verify the affine link (see, for instance, the analysis in Section 4 of \cite{Protter} for the 3/2 model).  
Further, we consider the class of stochastic volatility models based on polynomial processes, introduced in \cite{Cuchiero-PhD}
and  \cite{CKRT12}. The exponentially affine models by \cite{KMKV2011}, which exhibit an affine link between the variance swap rate and the spot variance, are included in the polynomial class, as a special case, see  Example 3.1 in \cite{CKRT12}.
In the polynomial framework, we prove the existence, in the presence of jumps, of a quadratic correction in the link between the theoretical variance swap rate and the spot variance.

In the financial market, traded variance swaps are actually written on the realized variance, that is the finite sum of squared log-returns sampled over a discrete grid. Instead, the corresponding theoretical pricing formulae use the continuous time approximation given by the quadratic variation of the log-price, in virtue of higher mathematical tractability. Thus, we also study the case where the theoretical variance swap rate, i.e., the expected future quadratic variation, is replaced by its empirical counterpart, namely the expected future realized variance. In this regard, we show that polynomial processes exhibit a quadratic link between the expected future realized variance and the (multidimensional) stochastic process characterizing the model. The pricing error related to this approximation has been investigated by \cite{BJ08}, who conclude that the approximation works quite well, based on simulated data obtained from four different models (the Black-Scholes model, the Heston stochastic volatility model, the Merton jump-diffusion model and the Bates stochastic volatility and jump model). 

Based on these results, our second contribution is testing, using OLS, if an affine or a quadratic link is satisfied by actual financial data, namely S\&P 500 daily data. This may allow us to determine which class of models, affine or polynomial, provides a better fit for empirical data. Clearly, such a test requires the availability of a the daily time-series of price, variance swap rate and spot variance observations. However, while S\&P 500 prices and variance swap rates, in the form of the (squared) VIX index (see \cite{CW08, VIXWhitePaper}), are quoted on the market, the spot variance is a latent process. Thus, the main hurdle impeding the testing of the affine/quadratic link is the latent nature of volatility process. To overcome this hurdle, the spot variance is estimated by means of the Fourier method proposed in \cite{MM09} and extended to jump-diffusions in \cite{CT15}.
The topic of the efficient estimation of the spot variance is relatively recent, unlike that of the efficient estimation of the integrated variance.
An early attempt goes through the following idea: first, the integrated variance is estimated over a local time window, relying on the realized variance formula; then, a localized spot variance estimate is obtained through a numerical derivative.
However, this differentiation-based estimation procedure gives rise to strong numerical instabilities, see
\cite{MyklandZhang2006,FosterNelson,ComteRenault}.
Further, this procedure requires the use of high-frequency prices to be efficient. However, it is well known that empirical high-frequency (tick-by-tick) prices are contaminated by microstructure noise, preventing the
realized variance from converging to the integrated variance of the price process.  The Fourier spot variance estimator allows to mitigate numerical instabilities, by relying on the integration of the price observations rather than on a differentiation procedure. Further, it results to be robust to different kinds of microstructure noise contaminations, provided that the highest frequency to be included in the Fourier series is chosen appropriately (see \cite{MaSan}).

The findings of our empirical tests are summarized as follows. First, we obtain overwhelming empirical evidence supporting the use of exponentially affine models in financial applications. Exponentially affine models imply the existence of an affine link between the variance swap rate and the spot variance, with strictly positive coefficients. The test of the affine link over the period 2006-2018 is coherent with this prediction, in that it yields statistically significant positive coefficients and an $R^2$ larger than $0.95$.
Instead, the test of the quadratic link between the variance swap rate and the spot variance over the period 2006-2018 yields a non-significant quadratic coefficient. 
%{\bf Therefore, as the polynomial model with an effective jumps contribution gives rise to a quadratic link, we can deduce that the extension to polynomial model over the whole data set should not be adopted.}
%%This result provides empirical evidence against the use of polynomial models with jumps, which imply the existence of a quadratic correction, while robustifying the finding related to the use of exponentially affine models.
%At the same time, t
This result may shed light on the negligibility of the discrete sampling effect affecting the variance-swap pricing formula. In fact, the absence of a significant quadratic coefficient confirms that the daily sampling used to compute the  VIX  index is enough to match the continuous-time approximation of the latter, i.e., the expected future quadratic variation.  This empirical finding, which is achieved in a non-parametric fashion, i.e., without assuming any parametric form for the price evolution, supports the numerical findings by   \cite{BJ08}.

The affine and quadratic tests are performed also on yearly subsamples, to investigate the sensitivity of the results to different economic scenarios. Test results on yearly subsamples are more nuanced. In particular, the intercept in the affine test is not significant in 2008 and 2011, two years characterized by extreme volatility spikes. This suggests that S\&P500 data in 2008 and 2011 are
consistent only with the broader assumption of an exponentially mean-reverting variance framework, which does not put any restrictions on the sign of the intercept (see, e.g., the rough-Bergomi model  in \cite{bayer2016pricing,jacquier2018vix}). Moreover, the quadratic test yields significant quadratic corrections in the years characterized by a relatively high number of price jumps. This findings support the use of polynomial models with jumps in periods when jumps are frequent. In general, our empirical analysis reveals that jumps play a non-negligible role, as we detect price-jumps in approximately $10\%$ of days of our 13-year sample. This result is in accordance with a large literature, see, e.g., \cite{BCC97, BarnShe2002, Bates96,  Eraker04}. Perhaps surprisingly, high-volatility periods and periods with a larger number of jumps do not necessarily coincide. For example, in 2007, 2010 and 2013, in spite of a relatively low  VIX  index, the number of days with jumps is relatively large.

%This result calls for the generalization of the Heston and Jacobi stochastic volatility models, see   \cite{AFP18}, to include jumps.

The paper is organized as follows. In Section \ref{sec:model} we describe the analytical framework of the paper, illustrating the exponentially affine model, the exponentially mean-reverting variance model and the polynomial model. In Section \ref{sec-jumps} we detail the spot variance estimation method and perform empirical tests to investigate if S\&P500 daily data over the period 2006-2018 are consistent with the affine or the quadratic link. Section \ref{Conclusions} concludes. The proofs are in the Appendix.

\section{Variance swap rate and model set-up}
\label{sec:model}

In this section we introduce the problem of the  variance swap valuation and investigate the types of models under which an affine link between the variance swap rate and the spot variance exists.

According to the fundamental theorem of asset pricing by   \cite{DS94},
the time evolution of the logarithm of the asset price follows a square-integrable semimartingale model, that is
\begin{equation}\label{eq:semi-mart}
X_t = A_t^{X} + M^{X}_t,
\end{equation}
where $M$ is a square-integrable martingale and $A$ is a finite-variation process on a filtered space $(\Omega,  \mathcal{F}, P)$. Being interested in the pricing problem, asset price dynamics are specified under a risk neutral measure along the paper. Moreover, in the paper we denote by $[X]_t$ the quadratic variation of the process $X$ up to time $t$. The semimartingale
hypothesis assures that the $[X]_t$ is finite for all times $t$ and coincides with the quadratic variation of the martingale $M^X$, if the finite-variation process $A$ has continuous paths.

A classical result proves that the quadratic variation can be obtained as the limit of the realized variance. More precisely, letting $\pi_m:=\{0=t_0< t_1< \ldots < t_m=\tau\}$ be a partition of
a generic interval $[0,\tau]$  and $|\pi_m|:= \displaystyle \sup_{k=1,\ldots, m}(t_k-t_{k-1})$ be the step of the partition, the realized variance is defined as
\begin{equation}
\label{RV}
RV^m_{[0,\tau]} = \sum_{k=1}^m (X_{t_k} -X_{t_{k-1}})^2.
\end{equation}
Then, the following convergence holds in probability
\begin{equation}
\label{eq:realized variance}
[X]_\tau= \lim_{|\pi_m|\rightarrow 0} RV^m_{[0,\tau]}.
\end{equation}
%We easily identify inside the parentheses the log-returns of the asset. By construction, quadratic volatility is a stochastic process itself
% excepting few basic models, for instance Black Scholes, where the quadratic variation is just proportional to the time.
A financial product, called {\sl variance swap}, was introduced to hedge volatility risk.
\begin{definition} (Variance Swap)
\label{varianceswap}
A variance swap is a financial derivative characterized by two legs, one paying the mean realized variance over an interval $[t,t+\tau]$,
the other paying a fixed amount, generally called the rate or strike.
Variance swap buyer pays the fix amount and receives the realized variance $RV^m_{[t,t+\tau]}$,
%generally computed using the square of daily log-returns of the  underlying, that is the realized variance reads
% $ \displaystyle RV_{[t,t+\tau]} = \sum_{k=1}^m (X_{t_k} -X_{t_{k-1}})^2$
with the convention that $t_k-t_{k-1}$ is one day, $t_0=t$ and $t_m =t+\tau$.
The strike $VS^\tau_t$ reads
\begin{equation}
\label{strike}
VS^\tau_t = \tau^{-1} {  E}\left[ RV^m_{[t,t+\tau]}  \, \vert \, {\cal F}_t \right].
\end{equation}
%where the expectation is computed under a risk neutral probability measure.
\end{definition}
Based on higher mathematical tractability, the finite-sample realized variance (\ref{RV}) is replaced, in the theoretical variance swap pricing formula, by its continuous-time approximation, the quadratic variation $[X]_\tau$. As a consequence, the strike of the variance swap (\ref{strike}), under the continuous-time limit, reads
\begin{equation}
\label{formula Variance-Swap}
VS^\tau_t = \tau^{-1} { E}\left[  [X]_{t+\tau} - [X]_t \, \vert \, {\cal F}_t \right].
\end{equation}
The simulation study by   \cite{BJ08}, based on four different models (the Black-Scholes model, the Heston stochastic volatility model, the Merton jump-diffusion model and the Bates stochastic volatility and jump model), suggests that the continuous-time approximation for the variance swap pricing formula works quite well.

A model-free pricing method, used to compute the VIX index (see  \cite{VIXWhitePaper}), has been also proposed by   \cite{carr2008robust}. This method exploits the fact that the variance swap can be perfectly statically replicated through vanilla Puts and Calls,  as pointed by the next result
(see   \cite{CW06} for the proof).

% The use of variance swaps rather than volatility swaps is justified since the first could be perfectly statically replicated through vanilla puts and calls, whereas a volatility swap requires dynamic hedging, as pointed by the next result see the proof in Carr and Wu \cite{CW06}.

\begin{proposition}(Variance Swap rate)
\label{prop-static-replication}
Assuming that the underlying asset price $X_t$ has continuous paths, then the variance swap can be statically replicated by a weighted position on
vanilla Puts and Calls, that reads
\begin{equation}\label{static-hedging-Variance-Swap}
VS^\tau_0 = \frac{2}{\tau} e^{r\tau} \left( \int_0^F  \frac{1}{K^2} P(K) {\rm d}K + \int_F^\infty  \frac{1}{K^2} C(K){\rm d}K  \right),
\end{equation}
where $F$, $\tau$ and $r$ denote, respectively, the forward of the underlying, the maturity and the risk-free interest rate, which is assumed to be constant.
The prices of the Call and Put options with strike $K$ and maturity $\tau$ are denoted, respectively, by $C(K)$ and $P(K)$.
\par\noindent
Moreover, in the presence of jumps in the price process $X_t$, the formula (\ref{static-hedging-Variance-Swap}) is subject to the correction $\epsilon_{\rm{J}}$, which depends only on the jump measure and reads
\begin{equation}
\epsilon_{\rm{J}} = -\frac{2}{\tau} \; {  E}\left[ \int_0^\tau \int \left( e^x -1-x-\frac{1}{2}x^2 \right) \nu({\rm d} \it t,  {\rm d }  x)\right],
\end{equation}
where $ \nu({\rm d}\it t, {\rm d}\it x)$ denotes the compensated Levy measure of the jump process.
\end{proposition}

As far as equity models are concerned, in this work we focus on a two-dimensional framework, where the first process is the logarithm of asset price as in (\ref{eq:semi-mart}) and the second, called variance process, is the variance of the martingale part in
(\ref{eq:semi-mart}) or a function of the latter. More precisely, in the rest of the paper we consider various model specifications within the following general class for the price evolution
\begin{equation}\label{eq-general-model}
    \begin{cases}   {\rm d}\it{X_t = \mu_t} {\rm d}\it t + \sqrt{V_t} {\rm d} \it B_t + {\rm d} \it J_t    \\ {\rm d} V_t = \alpha_t {\rm d} \it t + {\rm d}\it Z_t   \end{cases} 
\end{equation}

\noindent where $B$ is a Brownian motion, $J$ is a compensated jump process characterized by the Levy measure  $\nu$ and $Z$ is an integrable
stochastic process with zero mean.
Note that the process $Z$ is not required to be a semimartingale. This allows us to include also the fractional Brownian motion case, see, for instance, Section 7 of \cite{ELV07}.
The class of models (\ref{eq-general-model}) and its extension to multi-dimensional volatility processes are extremely large and include almost all stochastic volatility models commonly used in finance.

\subsection{Exponentially affine model}
\label{affine_sv_model}

With pricing and forecasting applications in mind, researchers focus on some subclasses of (\ref{eq-general-model}), which are able to capture equity stylized facts while still remaining parsimonious.
During the  last two decades, a large literature, started by   \cite{DufPanSin}, has focused on exponentially affine models, which are defined as follows, see Definition 2.1 in \cite{DufFilSch}.

\begin{definition}(Exponentially affine stochastic volatility model)
\label{DEF_affine_sv_model}
A Markov process $(X,V)$ is called affine if the characteristic function of the process has an exponential affine
dependence on the initial condition.
 
That is, for every $0\leq u <t$, there exists functions $(\psi^x_{(a,b)}(t,u), \psi^v_{(a,b)}(t,u), \phi_{(a,b)}(t,u))$ such that
$$
{  E}\left[ e^{aX_t+bV_t} \vert {\cal F}_u \right]  = \exp \, \{ X_u \psi^x_{(a,b)}(t - u) + V_u \psi^v_{(a,b)}(t - u) + \phi_{(a,b)}(t,u) \, \}.
$$
\end{definition}
Under natural financial hypotheses, we have $\psi^x_{(a,b)}(t-u) =1$. Moreover,   \cite{DufPanSin}
show that $\psi^v_{(a,b)}$ satisfies a generalised  first order non-linear differential equation of Riccati type
and $\phi_{(a,b)}$ is a primitive of a functional of $\psi^v_{(a,b)}$.

The most popular exponentially affine model, and the only one with continuous paths, is the model by  \cite{Heston},
which reads
 
\begin{equation}\label{eq-Heston}
\begin{cases}
{\rm d}X_t = \left(r- \frac{1}{2} V_t \right) {\rm d}t + \sqrt{V_t} {\rm d}B_t \\
{\rm d}V_t =\kappa (\theta -V_t) {\rm d}t + \sigma \sqrt{V_t} {\rm d}W_t 
\end{cases} 
\end{equation}

\noindent where $B$ and $W$ are correlated Brownian motions.
Moreover,  it is easy to verify that, under the Heston model, the variance swap strike (\ref{formula Variance-Swap}) has the following expression:
\begin{equation}
\label{Variance-Swap-Heston}
VS_t^\tau= \theta  + (V_t -\theta) \frac{1-e^{-\kappa \tau}}{\kappa \tau}.
\end{equation}

The class of exponentially affine models is wide, including also jumps processes, and has been extensively investigated, see for instance
\cite{Bates96, Benth11, BSS19, FM09, HKRS17, HX19, JMSZ19, KR2011}.
In this regard, we highlight the results by  \cite{KRST2011} and   \cite{CT13}, who show that exponentially affine processes are regular. Note that, in the exponentially affine framework, the variance process $V$ needs to be driven by a martingale $Z$ (see \eqref{eq-general-model}) with finite quadratic variation. Moreover, the drift process $\alpha$ and the Levy measure $\nu$ of the jump process $J$ in \eqref{eq-general-model} need to be affine with respect to the variance process $V$.

  \cite{KMKV2011} show that the affine link between the spot variance and the expected integrated variance holds for any exponentially affine stochastic volatility model. Their result is presented in the following proposition.
\begin{proposition}(Laplace transform of the quadratic variation)
\label{KAL}
Let $(X,V)$ be an exponential affine stochastic volatility model. Then, the triplet $(X_t,V_t,[X]_t)$ is a Markov exponentially affine process.
Moreover, the process $[X]_{t}$ has the following characteristic function
$$
{  E} \left[ e^{u [X]_{t+\tau} } \vert {\cal F}_t\right] = \exp \left\{u [X]_t + V_t \Psi^V_u(\tau)  + \Phi^V_u(\tau)   \right\},
$$
where $\Psi_u^V$ satisfies a couple of first order non-linear differential equations of Riccati type and $\Phi_u^V$ is a primitive of a functional of
$\Psi_u^V$. More precisely, using the parameter notation for the exponentially affine model introduced in Lemma 4.2 of 
\cite{KMKV2011}, they satisfy
\begin{eqnarray*}
\frac{\upartial \Psi^V_u}{\upartial t}(t) &=& \frac{1}{2}\gamma_1^{11} \left(\Psi^V_u(t)\right)^2 + \beta_1^1 \Psi^V_u(t) + \gamma_1^{22} u
 +\int_{\mathbb{R}^+\times \mathbb{R}} \left( e^{x_1\Psi^V_u(t) +ux_2^2} -1 - \Psi^V_u(t) h(x_1)  \right) \kappa_1({\rm d}x), \\
\Psi^V_u(0) &=& 0, \\
\Phi^V_u(t) &=& \int_0^t  \left[\beta_0^1 \Psi^V_u(s) +\gamma_0^{22} u + \int_{\mathbb{R}^+\times \mathbb{R}}
 \left( e^{x_1\Psi^V_u(s) +ux_2^2} -1 - \Psi^V_u(s) h(x_1)
\right) \kappa_0({\rm d}x) \right] {\rm d}s.
\end{eqnarray*}
Moreover, assuming that  $\int_0^\tau {  E}[V_t] {\rm d}t <\infty$, then
\begin{equation}\label{eq-general-affine-link}
VS^\tau_t =  V_t \Psi(\tau) + \Phi(\tau),
\end{equation}
where $\Psi(\tau)$ (respectively, $\Phi(\tau)$) is the partial derivative of $\Psi^V_u(\tau)$ (respectively, $\Phi^V_u(\tau)$) with respect to $u$, at $u=0$.
\end{proposition}
Note that this affine link is not satisfied by all stochastic volatility models with an explicit Laplace transform. For instance, it is not satisfied by the $3/2$ model of \cite{32} and by the $4/2$ model of \cite{G2016} , see, also, the analysis in Section 4.3
of \cite{Protter}.

In the following proposition we complete the result by   \cite{KMKV2011}, showing that the functions $\Psi(\tau)$ and $ \Phi(\tau)$ are strictly positive. This additional result is interesting in view of our empirical study of section \ref{sec-jumps}, where we test if
S\&P data are coherent with the exponential affine framework, based on the significance of the estimates of the coefficients in (\ref{eq-general-affine-link}).
The proof of this additional result crucially relies on the characterization of exponentially affine models by  \cite{F01}, who shows, under mild conditions (mainly the non-negativity of $V$), that the volatility process $V$ has to be a continuous-state branching processes with immigration in the exponentially affine framework.  Note that
the explicit stochastic differential equation satisfied by a generic continuous-state branching process with immigration is provided by  \cite{DL06} and   \cite{LM08}, who also detail the conditions to have a stationary distribution for the variance process. The existence of a stationary distribution is usually
considered as a natural property of the variance process.
\begin{proposition}
\label{prop-coeffcient-affine}
Let $(X,V)$ follow an exponentially affine stochastic volatility model and assume that the variance process $V$ admits a non-degenerate
 stationary distribution. Then $\Psi(t)>0$ and $\Phi(t)>0$ for all $t>0$.
\end{proposition}
Based on Proposition \ref{prop-coeffcient-affine}, the exponential affine framework could be rejected by empirical data if
any of the coefficient estimates is not  strictly positive.
In that event, it could be worth investigating the adequacy of the more general exponentially mean-reverting variance and polynomial frameworks, respectively detailed in Subsection \ref{exponential_sv_model} and \ref{POLY}.

\subsection{Exponentially mean-reverting variance model}
\label{exponential_sv_model}

In this subsection we introduce a more general subclass of the stochastic volatility models  included in (\ref{eq-general-model}), which we name {\sl exponentially mean-reverting variance models}. Moreover, we show that, under this paradigm, an affine relationship between the variance swap rate and the spot variance holds.
\begin{definition}(Exponentially mean-reverting model)
\label{DEF-expo-mean-rev}
The stochastic volatility model $(X,V)$ is an exponentially mean-reverting variance
model if $(X,V)$ satisfies
\begin{equation}
 \label{eq-general-mean-reverting}
   \begin{cases}
 {\rm d}X_t = \mu_t {\rm d}t + \sqrt{V_t} {\rm d}B_t + {\rm d}J_t\\{\rm d}V_t =  \kappa (\theta - V_t) {\rm d}t + {\rm d}Z_t
 \end{cases}
 \end{equation}
\noindent where $\kappa>0 $, the jump process $J$ is square-integrable and its Levy measure is affine in the volatility process.
%(ii) the drift $\alpha$ of the variance process is also affine, i.e., $\alpha_t = \kappa (\theta - V_t)$, with $\kappa>0$,
\end{definition}

A relevant example inside this class is the rough-Heston model
(see, e.g., \cite{ElEuch, el2019characteristic, gatheral2018volatility}).
In fact, we do not put any constraint on the process $Z$, except the fact that it has zero mean. Therefore, this class includes
not only exponentially affine processes but also processes driven by a fractional Brownian motion.

The next result shows that the expected quadratic variation of an exponentially mean-reverting variance
model is affine in the spot variance.
\begin{proposition}
\label{exponential_mean_reverting_Prop}
Let $(X,V)$ be an exponentially mean-reverting variance model, as defined in (\ref{eq-general-mean-reverting}). Then the expectation of the quadratic variation $[X]$ of the log-price is an affine function of the spot variance $V$, i.e., there exist deterministic functions $\Psi$ and $\Phi$ such that
\begin{equation}
{  E}\left[ [X]_\tau \right] = V_0 \Psi(\tau) + \Phi(\tau).
\end{equation}
\end{proposition}
Differently from the case of the exponentially affine framework, in this case the coefficients $ \Psi(\tau)$ and $\Phi(\tau)$ are not strictly positive.
A first example of a model satisfying Definition \ref{DEF-expo-mean-rev} but not Definition  \ref{DEF_affine_sv_model}   is the Hull-White
stochastic volatility model, see \cite{hull1987pricing}, under which the volatility is log-normal. In particular, the Hull-White model fits Definition \ref{DEF-expo-mean-rev} for
$\theta=0$ and $dZ_t = \xi  V_t dW_t$, where $W$ is a Brownian motion and $\xi>0$. A straightforward computation shows that the variance swap rate is linear with
respect to the spot volatility.  Moreover, note that this model only admits a degenerate steady-state distribution, namely a Dirac delta on zero.
A more interesting example is the rough-Bergomi volatility model, see \cite{bayer2016pricing}. This case could be seen as an extension of the Hull-White model, where the Brownian motion is replaced by a fractional Brownian motion with Hurst parameter smaller than $1/2$. The main mathematical difficulty inherent to
rough models is that the volatility is not a Markov process. This problem could be overcome by taking an infinite-dimensional point of view (see, e.g., \cite{jaber2019weak} and references therein).  The initial value of the variance process $V_0$ is then
replaced by a function $g_0$ that takes into account the initial conditions. Thus, under the infinite-dimensional viewpoint, the link between
the variance swap rate and the initial variance is the functional linear link between
the variance swap rate and the function $g_0$.  The particular case of the rough-Bergomi volatility model is studied in \cite{jacquier2018vix}, where a linear link is detailed.
Finally, note that it is not possible  to work with the function $g_0$ empirically, unless this function is assigned a parametric form.

In Section \ref{Affine_framework}, we show that empirical subsamples related to the years 2008 and 2011, where, respectively, the outbreak of the global financial crisis and the
Euro-zone debt crisis took place, exhibit a non-significant intercept parameter. This result can tilt the balance in favor of log-normal models like Hull-White
and rough-Bergomi during crisis periods.

%The main analysis will be to study S\&P data in order to infer if they are coherent with a stochastic volatility model of exponential mean reverting framework or not.
%In order to discern, we exploit the affine link between variance swap rates and spot variance.
%According with Proposition \ref{prop-coeffcient-affine}, exponential affine framework could be rejected if the two coefficients are not both
%strictly positive. In that case, the more general case of polynomial model that will be studied in the next subsection, could be investigated.
%Moreover, we will also test if a second order correction perform better than an affine relation, according with Proposition \ref{prop-polynomial}.
%If the coefficient of the quadratic term is not significative, we could deduce that the discrete sampling is frequent enough to reach
%the continuous limit.

\subsection{Polynomial model}
\label{POLY}

In this section we consider the class of stochastic volatility models based on polynomial processes, introduced in   \cite{Cuchiero-PhD}
and   \cite{CKRT12}. As pointed out in  \cite{CKRT12}, exponentially affine processes are polynomial processes.
Moreover, under suitable restrictions, the polynomial class could be considered as a sub-class of
\eqref{eq-general-model}, see   \cite{C18}.

Let ${\cal P}_k$ denote the vector space of polynomials up to degree $k$. In the bi-dimensional case, we have the following
definition of a polynomial process.

\begin{definition}(Polynomial Process)
\label{DEF_polynomial}
A time-homogeneous Markov process $(X,V)$ is said $m$-polynomial, if, for all $k\in \{0,\ldots, m\}$,
$f \in \mathcal{P}_k$, $(x,v)$ in the state space and $t > 0$, it holds that
\begin{equation}
(x,v) \rightarrow {  E}^{(x,v)}\left[ f(X_t,V_t) \right] \in {\cal P}_k,
\end{equation}
where, for any $0\leq u<t$, we adopt the standard notation ${ E}^{(x,v)}\left[ f(X_t,V_t) \right] = {  E}[f(X_t,V_t)| X_u=x, V_u=v]$.
Also, the semigroup is assumed to be strongly continuous. Moreover, if $(X,V)$ is $m$-polynomial for all $m\in {\mathbb{R}}$, then $(X,V)$ is said
polynomial.
\end{definition}

A relevant non-affine example in this class is the Jacobi stochastic volatility model, see \cite{AFP18}. Other examples and
applications of polynomial process could be found in \cite{AF16, CFP17, Cuchiero-PhD, C18,  CLSF18,  FGM16}.

The following proposition allows us to investigate the existence of a quadratic correction in the link between
theoretical variance swap rates and spot variance in the polynomial framework.
\begin{proposition}
\label{prop-polynomial}
Let $(X,V)$ be a $2$-polynomial process describing a stochastic volatility model, then the expected quadratic variation of $X$ belongs to ${\cal P}_2$
in $(x,v)$. Moreover, if $(X,V)$ has continuous paths, then the expected quadratic variation of $X$
is affine in $(x,v)$.
%Moreover restricting $(X,V)$ to be a $2$-polynomial process describing a stochastic volatility model then the discrete realized variance
%of $X$ is quadratic in $(X,V)$.
\end{proposition}
This result suggests that the presence of a statistically significant quadratic correction could be explained by the presence of jumps in the underlying. In fact, the empirical analysis in Section \ref{Empirical} seems to support this finding.  In particular, in Section \ref{Quadratic_framework}, we point out that a quadratic coefficient is statistically significant in the years with a higher frequency of price jumps.

We conclude the section by discussing the effects of discrete sampling on the functional form of the variance swap rate. Indeed, the actual  price of traded variance swaps relies on the computation of the realized variance in place of its asymptotic approximation, given by the quadratic variation (see Definition \ref{varianceswap}). In this regard, the following result holds.

\begin{proposition}
\label{sampling}
%Recalling that the price of the variance swap relies on the computation of the realized variance (see Definition \ref{varianceswap}), it is possible to
% obtain the following result for a $2$-polynomial process.
%show that,
If $(X,V)$ is a $2$-polynomial process describing a stochastic volatility model, then the expected realized variance of $X$ belongs to ${\cal P}_2$.
%is quadratic in $(X,V)$.
\end{proposition}
Based on Proposition \ref{sampling}, the variance swap rate for a polynomial stochastic volatility model is at most quadratic in $(X,V)$, that is there exist coefficients $p_{i,j}(\cdot), \, i,j=0,1,2$, such that
\begin{equation}
\label{eq-general-quadratic-link}
 VS^\tau_t =  p_{0,0}(\tau) + p_{1,0}(\tau)\; X_t + p_{0,1}(\tau) \; V_t + p_{2,0}(\tau) \;  X^2_t + p_{1,1}(\tau) \; X_t\, V_t + p_{0,2}(\tau) \;  V^2_t \, .
\end{equation}
This result is interesting in that it may help collect empirical evidence supporting  the result by   \cite{BJ08}. The authors show, for some well known models, that the expected quadratic variation provides an efficient approximation of the actual  VIX  index, whose computation is based on a daily sampling scheme (see \cite{CW06, VIXWhitePaper}). In other words, non-significant  estimates of the quadratic coefficients in (\ref{eq-general-quadratic-link}) may represent empirical evidence that the continuous-time approximation works well enough.

Finally, note that, based on Proposition \ref{prop-polynomial}, the expression (\ref{eq-general-quadratic-link}) is also implied by the assumption that the data-generating process is a polynomial stochastic volatility model with jumps. Section \ref{Empirical} analyzes if a second order correction fits S\&P500 data better than the affine link implied by the affine framework (\ref{eq-general-affine-link}).

\section{Empirical study}
\label{sec-jumps}

In this Section we perform an empirical study to investigate if S\&P500 daily data over the period $2006-2018$ are consistent with the affine framework (see Paragraph \ref{Affine_framework}) or the polynomial framework (see Paragraphs \ref{Quadratic_framework} and \ref{Polynomial_framework}), based on the statistical significance of the estimates of the coefficients $p_{i,j}(\cdot), i,j=0,1,2$, in (\ref{eq-general-quadratic-link}).
To perform this study, we use the daily series of variance swap rate and log-price observations, plus a daily series of estimates of the unobservable spot variance. Accordingly, this Section begins with the description of the dataset used for the empirical exercise, while Section \ref{spotvolest} describes the method employed to reconstruct the spot variance path on a daily grid from the series of high-frequency prices in the dataset.

\begin{figure} 
\includegraphics[width=\linewidth]{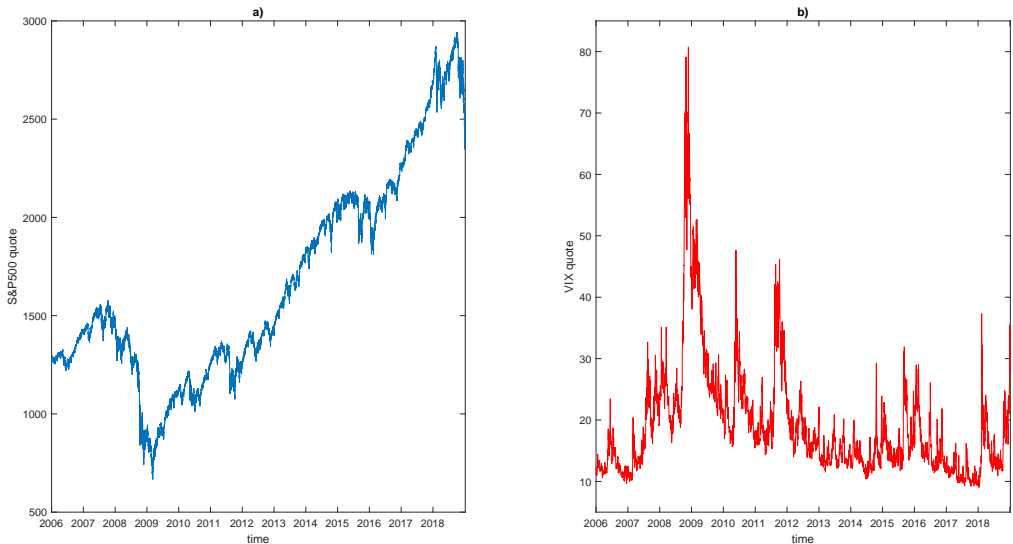}
\caption{ Panel a): 1-minute S\&P500 trade prices over the period 2006-2018; panel b):  Opening quotes of the VIX index over the period 2006-2018.} \label{dataset}
\end{figure}

The dataset, ranging over the period 2006-2018, is composed of the series of S\&P500 trade prices, recorded at the $1$-minute sampling frequency (see panel a) in Figure \ref{dataset}), and the series of VIX index values, recorded at the beginning of each trading day (see panel b) in Figure \ref{dataset}).
The period 2006-2018 encompasses a number of volatility peaks, corresponding to historical financial events such as the global financial crisis of 2008, the flash-crash of May 2010, the Eurozone debt crisis of 2011, the Brexit events of 2016 and the US-China `trade war' of 2018. For a detailed description of the events that have affected the US stock market since the 1990s, see \cite{HX19}.

%Note that the squared VIX index is an accurate proxy of the rate of the variance swap with maturity one month. In \cite{CW06} it is shown that, under very general assumptions, the one-month variance swap rate coincides with the VIX index squared up to an error term due to the presence of jumps in the price process (see Proposition \ref{prop-static-replication}). However, since this error term is also affine (resp. polynomial) under the assumption that the data-generating process is an affine (resp. a polynomial) stochastic volatility model, it can be encompassed in our empirical study, where jumps play a role, as discussed in Section {sec:model}.
%The correction $\epsilon^J$ detailed in Proposition \ref{prop-static-replication} depends on the Levy measure of underlying jump process.
%Then it is affine under exponential affine, see \cite{DufPanSin}, and exponential mean reverting frameworks according to Definition \ref{DEF-expo-mean-rev}.
%{\bf SECONDO ME TIRARE IN BALLO IL J NON SERVE E NON SI CAPISCE PIU CHE FORMULA USIAMO}

\subsection{Spot variance estimation}
\label{spotvolest}

The latent spot variance at the beginning of each trading day is reconstructed from $1$-minute prices using the Fourier methodology, according to \cite{MM09} and \cite{CT15}. More precisely, we first detect the presence of jumps in our sample data and, secondly, we use the
the estimator by \cite{CT15}, with sparse sampling, in days where jumps are detected. In the other days, the estimator by \cite{MM09} is employed with the entire high frequency data set.
For the reader convenience, we briefly recall the definition and main properties of these spot volatility estimators.

The Fourier estimator by \cite{MM09} is defined as follows. Let $[0,2\pi]$ be the time horizon and consider the time grid
$\{ 0=t_{0}< \ldots < t_m=2\pi\}$.
For any integer $k$, $|k|\leq 2N$, define the discrete Fourier transform
\begin{equation}\label{FOUMULT1}
c_k({\rm d}X_{m}):= {1\over {2\pi}} \sum_{j=1}^{m} e^{-{\rm i}kt_{j-1}}(X_{t_{j}}-X_{t_{j-1}}).
\end{equation}
Then, for any integer $k$, $|k|\leq N$, consider the following {\sl convolution} formula
\begin{equation}
\label{coefSTIMvol}
c_k(V^m_{N}):= {2\pi \over {2N+1}} \sum_{|h|\leq N} c_h({\rm d}X_m) c_{k-h}({\rm d}X_m).
\end{equation}
Formula (\ref{coefSTIMvol}) contains the identity relating the Fourier transform of the log-price process $X_t$ to the Fourier transform of
the variance $V_t$.
By (\ref{coefSTIMvol}) we gather all the Fourier coefficients of the variance function by means of the Fourier transform of the
log-returns. Then, the reconstruction of the variance function $V_t$ from its Fourier coefficients is obtained through the Fourier-F{e}j\'{e}r summation, i.e., the {\sl Fourier estimator of the spot variance} is defined as follows: for any $t\in (0,2\pi)$,
\begin{equation}
\label{FouSpotEst}
\widehat V^m_{N,M}(t)= \sum_{|k|< M} \left(1- {|k|\over M}\right) c_k(V^m_{N}) \ {\rm e}^{{\rm i}k t}.
\end{equation}
We note that the definition of the estimator $\widehat V^m_{N,M}(t)$ depends on three parameters, the number of data $m$ and the two {\sl cutting frequencies} $N,M$. An appropriate choice of the cutting frequencies is needed to filter out the microstructure noise effects. In fact, on one side
the estimation of the instantaneous volatility benefits from the availability of a large amount of data, at least from a statistical point of view. On the other side, high-frequency data are affected by microstructure noise effects deriving from, e.g., bid-ask bounces, infrequent
trading and price discreteness. Therefore, it is necessary to employ volatility estimators which are able to filter out microstructure noise contaminations. The estimator of the spot variance $V_t$ by means of the Fourier method has been designed to this aim, and is robust to the presence of different types of noise contaminations in the price process, see \cite{MaSan}.

The Fourier method to estimate the spot variance has been extended to the case where jumps are present in the price process by \cite{CT15}. The procedure has two steps. First, an estimate of the Fourier coefficients of a continuous invertible function $\varrho(V)$ of
the instantaneous variance is obtained. The estimator of the $k$-th Fourier coefficient takes the form
\begin{equation}
\label{FouCoefCT}
\sum_{j=1}^{m} {1\over m} \, {\rm e}^{-{\rm i} k t_{j-1}} g(\sqrt{m} (X_{t_{j}}-X_{t_{j-1}})),
\end{equation}
where the function $g$ can assume different specifications. We will consider $\varrho_g(V_t)= {\rm e}^{-{V_t/ 2}}$, that is we choose here $g(x)={\rm cos} \ x$.
Second, we invoke the Fourier-Fej\'{e}r inversion formula as in (\ref{FouSpotEst}) to reconstruct the path of the process
$\varrho(V)$ as follows:
\begin{equation}
\label{FourierJC}
\frac{1}{2\pi}\sum_{j=1}^{m}\frac{1}{m}F_M (t-t_{j-1}) \ g(\sqrt{m} (X_{t_{j}}-X_{t_{j-1}})),
\end{equation}
where $F_M(x)$ is the Fej\'{e}r kernel. Note that also (\ref{FouSpotEst}) can be re-written by means of $F_M(x)$, see \cite{MaRec}.
Finally, this  is translated into an estimator of the spot variance
by inverting the function $\varrho(\cdot)$. The obtained estimator of the instantaneous variance is consistent and asymptotically efficient in the absence of microstructure noise.

\begin{figure} 
\includegraphics[width=\linewidth]{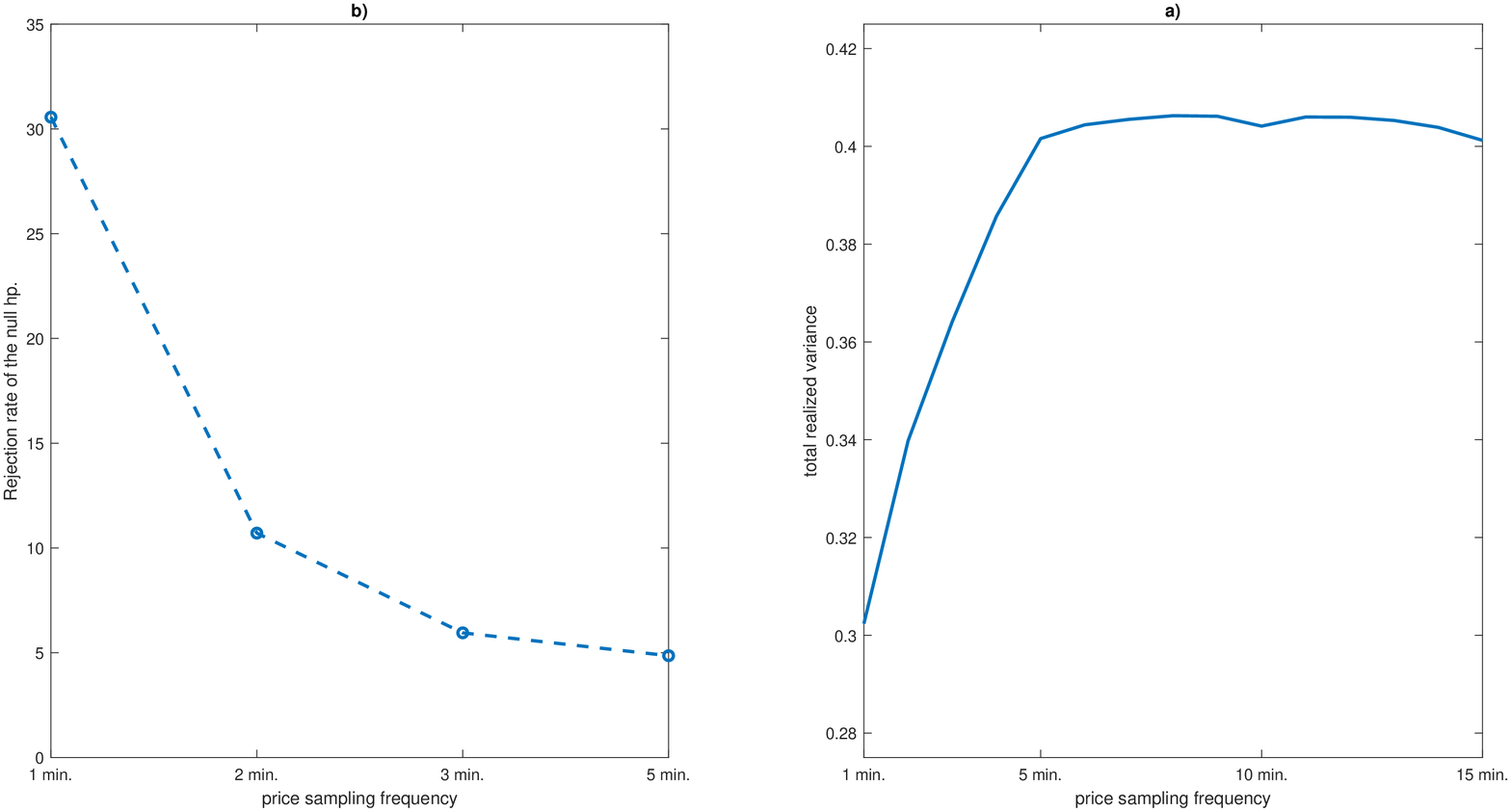}
\caption{Panel a): rejection rate of the null hypothesis of the noise-detection test by \cite{AX19}, performed on daily subsamples of S\&P500 prices over the period 2006-2018, for different sampling frequencies; panel b): volatility signature plot, i.e., total S\&P500 realized variance for the period 2006-2018 as a function of the price sampling frequency.} \label{sign}
\end{figure}

In order to assess whether the characteristics of the S\&P500 $1$-minute prices data require either the use of the jump-robust Fourier estimator of spot volatility or not,
%are compatible with the assumptions under which a given estimator is consistent. Accordingly, prior to obtaining spot volatility estimates from our sample of $S\&P500$ 1-minute prices,
we perform the following tests.
First, we split the sample into daily subsamples and apply the test by \cite{AX19} for the null hypothesis that the price is an
%, potentially discontinuous,
It\^o semimartingale.
%without noise contaminations.
Test results at the $95\%$ confidence level, illustrated in the Figure \ref{sign}, panel a), show, consistently with the literature \cite{ABDE01}, that the impact of microstructure noise on prices can be considered negligible at the $5$-minute frequency. This finding is consistent with the behavior of the volatility signature (Figure \ref{sign}, panel b)), which shows that the total Realized Variance of the sample stabilizes, around $0.4$, from the $5$-minute frequency and downwards.
%\begin{table}[H]
%\begin{center}
%\begin{tabular}{ c |  c | c | c | c     }
%Price sampling frequency  & 1 min. & 2 min.   & 3 min. & 5 min.   \\   \hline
% Rejection rate of the null hypothesis  & 0.3056 & 0.1071& 0.0595  &0.0486   \\
%\end{tabular}
%\end{center}
%\caption{Results of the noise detection test by \cite{AX19}: .}
%\end{table}

\begin{figure} 
\includegraphics[width=\linewidth]{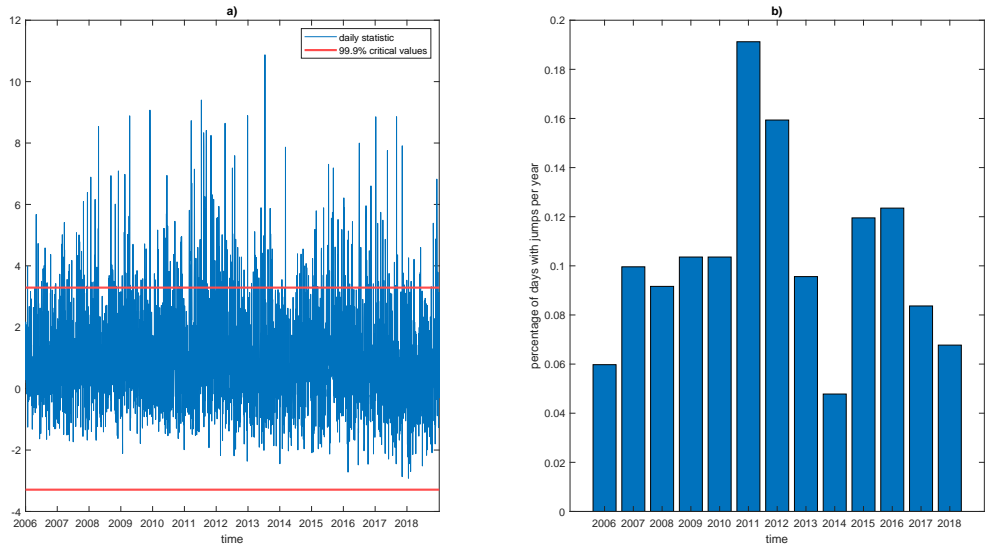}
\caption{Panel a): values of the statistic of the jump-detection test by \cite{cpr} computed over the period 2006-2018; panel b): ensuing percentage of days with jumps per year.}
\label{boh}
\end{figure}

\begin{figure} 
\includegraphics[width=\linewidth]{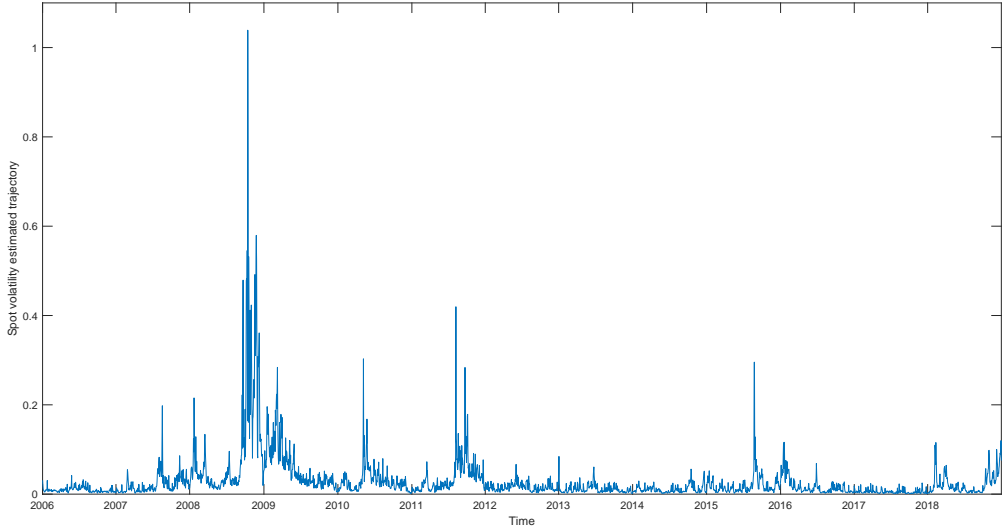}
\caption{Estimated S\&P500 spot variance trajectory on a daily grid over the period 2006-2018.}
\label{SPOT}
\end{figure}

Secondly, after downsampling the log-price series at the $5$-minute frequency, we apply the jump detection test by \cite{cpr} for the null hypothesis that the price is a continuous semimartingale.
%(without noise contaminations).
Test results at the $99.9\%$ confidence level show that jumps are detected in $10.35\%$ of the daily subsamples over the period 2006-2018. Figure \ref{boh} shows the values of the  test statistic computed from daily subsamples (panel a)) and the ensuing  percentage of days with jumps per year (panel b)). %Note that the latter displays a peak in 2011, reaching about 20\%, and, in general, is higher
%in the years 2011-12 and 2015-16, corresponding to two phases of financial turmoil
The percentage of jumps detected per year is compatible with the empirical results in \cite{cpr}.
Based on the results of the two tests, in order to obtain spot variance estimates, we proceed as follows. On the consecutive days in which the hypothesis of absence of jumps is not rejected (amounting approximately to $90\%$ of the sample), the Fourier estimator (\ref{FouSpotEst}) is applied with all prices recorded at $1$-minute frequency. Instead, for the sparse days in which jumps are detected (amounting approximately $10\%$ of the sample)
we use spot variance estimates obtain through the Fourier estimator (\ref{FourierJC}), applied to sparsely sampled 5-minute prices.
%\footnote{As mentioned, we detect the presence of jumps in 10.35\% of daily subsamples over the period 2006-2018, corresponding to 338 out of 3267 days. Of these 338 days, 56 are immediately followed by another day with jumps, 9 are followed by two consecutive days with jumps, 4  are followed by three consecutive days with jumps, and only one is followed by four consecutive days with jumps. This implies that 261 days with jumps are isolated, i.e. are in between two days in which the hypothesis of absence of jumps is not rejected.},
%\footnote{To clarify our procedure, consider the following example. Assume, e.g., that a 20-day sample were available, and that the presence of jumps were detected in days 2,3 and 9. In such a case, we would split our sample into 5 subsamples: the first, coinciding with day 1; the second, encompassing days 2 and 3; the third, ranging from day 4 to day 8; the fourth, coinciding with day 9; the last, ranging from day 10 to 20. Then, we would use the estimator by Malliavin and Mancino for the first, third and fifth subsamples, and the estimator by Cuchiero and Teichmann for the second and fourth subsamples.}.
In the case of the estimator (\ref{FouSpotEst}), the cutting frequencies have been selected as $N=m^{2/3}/2$ and $M=m^{2/3}/(16 \pi)$, according to \cite{MaRec}, who find these cutting frequencies to be optimal in the presence of different types of noise and noise intensities. For the estimator (\ref{FourierJC}),
instead, the frequency $M$ is selected as $M= (m/4)^{2/3}$, in accordance to \cite{CT15}.
%such that the spot volatility is estimated on a grid of mesh size equal to $30$ minutes \textcolor{red}{METTERE $M$ Questa scelta di M andrebbe giustificata}.
%Note that, in the case of the estimator (\ref{FourierJC}), the selection of M dictates the mesh size of the spot volatility estimation grid; therefore, in days with jumps, we retain only the spot volatility estimate at the beginning of the day, as only that estimate is relevant for our empirical study. Instead, in the case of the estimator (\ref{FouSpotEst}), we directly obtain spot volatility estimates at the beginning of each trading day, as the estimation grid is independent of N and M.
The resulting spot variance estimates at the beginning of each trading day are shown in Figure \ref{SPOT}.

\subsection{Empirical test results}
\label{Empirical}

We now focus on testing the empirical link between the rate of the variance swap with time to maturity equal to one month, i.e., the VIX index squared, and the couple spot variance - log return.
If estimates of the unobservable spot variance are available, equation (\ref{eq-general-quadratic-link})
can be rewritten, in the case of the S\&P500 index, as follows.
Let $L=3267$ denote the number of days in our sample and let $\tau=1/12$.  Then, for $t_i=\frac{i-1}{252}$, $i=1,2,...,L$, we write
\begin{equation}\label{feasiblepoly2}
{\rm VIX}^2_{t_i}  =  p_{0,0}(\tau) +  p_{1,0}(\tau) \, X_{t_i} + p_{0,1}(\tau)  \, \widehat V^m_{{t_i}}  + p_{2,0}(\tau)  \,  X^2_{t_i} + p_{1,1}(\tau)  \, X_{t_i}\,  \widehat V^m_{{t_i}}  +  p_{0,2}(\tau)  \, (\widehat V^m_{{t_i}})^2,
\end{equation}
where:
\begin{itemize}
\item[-] $ {\rm VIX}_{t_i}$ denotes the opening quote of the VIX index on $i$-th day;
\item[-] $X_{t_i}$ denotes the opening log-price of the S\&P500 index on $i$-th day;
%\item[-] $\epsilon_{t_i}^J$ denotes the difference between $(VIX)^2_{t_i} $ and $VS_{t_i}^{1/12}$ due to the presence of jumps in the price process, see  Proposition \ref{prop-static-replication}; recall that, if the price process is continuous, then $\epsilon_{t_i}^J = 0$;
%\item[-] $\epsilon_t^D$ denotes the difference between $(VIX)^2_t$ and $VS_t^{\tau}$ due the  discretization of the integral in the VIX formula  (see, again, \cite{CarrWu});
\item[-] $\widehat V^m_{{t_i}}$
%:= V_{t_i} + \eta_{t_i,n}$
denotes the estimated spot variance at the beginning of the $i$-th day, obtained from a sample of size $m$ using the estimator (\ref{FouSpotEst}) or (\ref{FourierJC}).
%since the order of magnitude of the finite-sample error, $\eta_{t_i,n}$, is negligible for large values of $n$, we exclude the term $\eta_{t_i,n}$ from the equation.
 %\item[-] $\eta_{t_i,n}$ denotes the finite-sample error related to estimation of $V_{t_i}$ from a sample of size $n$; if the estimator of $V_{t_i}$ is consistent, then $\eta_{{t_i},n} \to 0$ in probability as  $n \to \infty$ in the infill-asymptotics sense.
\end{itemize}
Some comments are needed.
First, based on the results of Propositions \ref{KAL} and \ref{prop-polynomial}, the presence of jumps does not spoil the affine/polynomial structure, thus the regression coefficients $p_{i,j}(\tau), \ i,j=0,1,2$, include the potential contribution of jumps. In the following, we drop the argument $\tau$ from $p_{i,j}(\tau)$ as we always consider monthly coefficients. Secondly, the consistency of the spot variance estimators (\ref{FouSpotEst}) and (\ref{FourierJC}) allows us to neglect the finite-sample error related to estimation of $V^m_{t_i}$.

We aim at testing the significance of the estimates of the coefficients in equation (\ref{feasiblepoly2}) within three progressively broader frameworks: the affine framework, introduced in Definition \ref{DEF_affine_sv_model} and extended in Definition \ref{DEF-expo-mean-rev} (hereafter, {\sl affine framework}); the polynomial framework of Definition \ref{DEF_polynomial}, where the variance swap rate is first assumed to be a quadratic function of $V$ only (hereafter, {\sl quadratic framework}) and then is  assumed to be a polynomial function of the couple $(X,V)$ (hereafter, {\sl polynomial framework}).

\subsubsection{Affine framework}
\label{Affine_framework}

In this paragraph we consider the exponentially affine and exponentially mean-reverting variance models, which both imply the existence of an affine relationship between the variance swap rate and the spot variance. Note that
%, while in the polynomial framework all coefficients in the equation (\ref{feasiblepoly2}) are non-zero,
in the affine framework the equation (\ref{feasiblepoly2}) reduces to
%quadro esponenziale affine e mean reverting a seconda che coefficientezero.
\begin{equation}
\label{model1}
{\rm VIX}_{t_i}^2  =  p_{0,0}  + p_{0,1}  \, \widehat V^m_{{t_i}}.
\end{equation}
Recall that the main discriminant factor between the exponentially affine model and its extension to the exponentially mean-reverting variance class is that
the former implies the coefficients $p_{0,0}$ and $p_{0,1}$ in equation (\ref{model1}) are strictly positive. Thus, we are not only interested in testing if the affine dependence between the variance swap rate and the spot variance is satisfied by empirical data, but also in verifying if both parameter estimates are significantly different from zero, as this would allow us to accept or reject the exponentially affine framework.

The coefficients in (\ref{model1}) are estimated using OLS.  Note that ${\rm VIX}^2_{t_i}$ is scaled as $({{\rm VIX}_{t_i}}/{100})^2/({30}/{365})$, in order to obtain properly normalized coefficient estimates.
In order to avoid performing a spurious regression (see \cite{Granger}), we first test for the null hypothesis of the presence of a unit root in the VIX squared series and in the  spot volatility estimates series, using the Augmented Dickey-Fuller test (see \cite{DFtest}). For both series, test results at the $99\%$ confidence level reject the null hypothesis. Thus, the two series are assumed to be stationary in the rest of the analysis.
The results of the OLS estimation are overwhelming: we obtain an $R^2$ larger than $0.95$ and significant coefficients estimates, as shown in Table \ref{Table_1}. In particular, the fact that both coefficient estimates are significant and positive suggests that an exponentially affine framework is a suitable fit for the S\&P500 data over the period 2006-2018. Note that the regression standard errors  have been computed using the Newey-West methodology (see \cite{NeweyWest}), to account for the presence of heteroskedasticity and autocorrelations in the residuals.

\begin{table}[h!]
\begin{center}
\caption{Affine framework (\ref{model1}): estimation from S\&P500 data over the period 2006-2018 \\(p-values less than $10^{-4}$ are reported as zero).}
 \label{Table_1}
 
 \vspace{0.25cm}
\begin{tabular}{ c ||  c   c   c  c   c ||  c      }
{framework} & {coeff.} & {estimate}   & {std. err.} & {t stat.} &  {p value}  & ${R}^2$  \\ \hline
affine & $ p_{0,0} $ &0.0011 & 0.0003 & 4.1312 & 0  & 0.9582\\
         &  $ p_{0,1} $ & 0.0801 &  0.0106  &  7.5512 & 0\\
  \end{tabular}
  
 \end{center}
 \end{table}

A natural question that arises is whether the coefficients in (\ref{model1}) are sensitive to events of distress, such as  the global financial crisis of 2008, or are stable over time, instead. To investigate this aspect, the coefficients of (\ref{model1}) are estimated on yearly subsamples. %to verify whether this framework is a suitable fit under different economic situations.
%usando la 2.5 il caso esponenziale affine i coefficienti entrambi significativi
%quindi preso tutto il set non si puo' rifiutare il espon affine percheè aentrambi esponenziaili affibi con salti non rifiutare
%nei sottoanni invece i coeeficienti non è piu evidente che l'intercetta sia sempre signifiatii, in parrticolare quando mercati sotto stres 2008 2011, l'intercetta non è significativi, dei due resti nell'affine, resta comunque mean reverting, che include il caso exponenziale affine (mean reverting piu larga)
%ma esclude i modelli 3/2 etc.
Estimation results  are detailed in Table  \ref{Table_5} and offer interesting insights.
\begin{table}[h!!]
\begin{center}

\caption{Affine framework (\ref{model1}): estimation from S\&P500 data over yearly subsamples\\ (p-values less than $10^{-4}$ are reported as zero).  }
\label{Table_5}

\vspace{0.25cm}
\begin{tabular}{  c ||  c   c   c   c   c | | c      }
 \multicolumn{7}{ c }{ {Affine framework}} \\
 \hline

  {year} & {coeff.} & {estimate}   & {std. err.} & {t stat.} & {p value}  & ${R}^2$  \\

\hline
   2006 & $p_{0,0} $ &0.0009 & 0.0001 & 10.5224 & 0  &    0.4405 \\
          &  $ p_{0,1} $ & 0.0394 &  0.0144  &  2.7279 & 0.0064\\
   \hline
   2007 & $p_{0,0} $ &0.0012 & 0.0003 & 3.6437 & 0.0003  &      0.6587\\
          &  $ p_{0,1} $ & 0.0714 &  0.0171  &  4.1778 & 0\\
  \hline
 2008 & $p_{0,0} $ &0.0017 & 0.0013 & 1.3576  & 0.1746  & 0.9795\\
          &  $p_{0,1} $ & 0.0776 &  0.0147  &  5.2930 & 0\\
  \hline
 2009 & $p_{0,0} $ &0.0026 & 0.0007 & 3.8468 & 0.0001  & 0.8630\\
          &  $p_{0,1} $ & 0.1095 &  0.0134  &   8.1877 & 0\\
  \hline
   2010 & $p_{0,0} $ &0.0021 & 0.0007 & 3.1524 & 0.0016  &     0.8235\\
          &  $p_{0,1} $ & 0.0779 &  0.0261  &  2.9862 & 0.0028\\
  \hline
   2011 & $p_{0,0} $ &0.0012 & 0.0011 & 1.0170 & 0.3092  &     0.9110\\
          &  $p_{0,1} $ & 0.1194 &  0.0344  &  3.4696 & 0.0005\\
  \hline
   2012 & $p_{0,0} $ &0.0019 & 0.0002 & 12.4158 & 0  &     0.3193\\
          &  $ p_{0,1} $ & 0.0458 &  0.0103  &  4.4609 & 0\\
  \hline
   2013 & $ p_{0,0} $ &0.0013 & 0.0001 & 17.2232 & 0  &     0.4952\\
          &  $ p_{0,1} $ & 0.0292 &  0.0069  &  4.2261 & 0\\
  \hline
   2014 & $ p_{0,0} $ &0.0012 & 0.0001 & 8.0686 & 0  &     0.5410\\
         &  $ p_{0,1} $ & 0.0405 &  0.0179  &  2.2641 & 0.0236\\
  \hline
   2015 & $p_{0,0} $ &0.0012 & 0.0005 & 2.4669 & 0.0136   &    0.8869\\
         &  $ p_{0,1} $ & 0.0579 &  0.0276  &  2.0938 & 0.0363\\ \hline

   2016 & $ p_{0,0} $ &0.0012 & 0.0001 & 15.9943 & 0  & 0.8168\\
         &  $ p_{0,1} $ & 0.0499 &  0.0048  & 10.3033  & 0\\
  \hline
   2017 & $ p_{0,0}  $ &0.0009 & 0.0001 & 21.6109 & 0  &  0.2484\\
          &  $ p_{0,1} $ & 0.0204 &  0.0050  &  4.0758 & 0\\
  \hline
   2018 & $p_{0,0} $ &0.0010 & 0.0001 & 10.0552 & 0  & 0.9122 \\
         &  $p_{0,1} $ & 0.0627 &  0.0051  &  12.1783 & 0\\
\end{tabular}

\end{center}
\end{table}
In periods of distress, like 2008, when the global financial crisis broke out, or 2011, when the financial turmoil related to sovereign debt crisis in the Euro area took place,
the intercept estimates are not significant at the 95\% confidence level. Thus, based on Proposition \ref{prop-coeffcient-affine}, S\&P500 data in 2008 and 2011 look consistent only with the broader assumption of an exponentially mean-reverting variance data-generating process, which poses no restrictions on the sign of the coefficients. In other words,
results on yearly subsamples tilt the balance in favor of the use, during crisis periods, of models that imply a linear relationship between the variance swap rate and the spot variance, such as the model  by \cite{hull1987pricing}
and the rough-Bergomi by \cite{bayer2016pricing, jacquier2018vix} model (see the discussion in Section \ref{exponential_sv_model}).
Finally, note that the empirical analysis suggests that the drift of the variance process is affine in the variance itself. As a consequence, models with stronger mean reversion, e.g., the $3/2$ model, are not coherent with our empirical findings. 

\subsubsection{Quadratic framework}
\label{Quadratic_framework}

In this paragraph we extend the analysis to take into account a possible quadratic link between variance swap rates and spot variance.
Through this section, the term {\sl quadratic} is used to refer to polynomial models that are not exponentially affine,
where no risk of confusion exists.
According to Propositions \ref{prop-polynomial} and \ref{sampling}, when polynomial models with jumps are considered and/or discrete-sampling effects can not be neglected, the variance swap rate is a bivariate polynomial in $X$ and $V$. 
Based on these results, the rest of our empirical study is aimed to investigating  the possible existence of a quadratic link between the variance swap rate and the spot variance and/or the log-price. In particular, the analysis is split into two parts. In this paragraph, devoted to what we have called the {\sl quadratic framework}, we check if a quadratic form with respect to the spot variance fits the data better than an affine form. In the next paragraph we will deal with the general form as in \eqref{feasiblepoly2}. In the quadratic framework, equation (\ref{feasiblepoly2}) reads
\begin{equation}\label{model2}
{\rm VIX}^2_{t_i}  =  p_{0,0}  + p_{0,1} \, \widehat V^m_{{t_i}}  + p_{0,2} \, \left(\widehat V^m_{{t_i}}\right)^2.
\end{equation}
%where $r_{{t_i},n} := \tilde p_{0,1}   \eta_{{t_i},n} + 2\tilde p_{0,2}  \hat V_t \eta_{{t_i},n} +\tilde p_{0,2}  \eta_{{t_i},n}^2$.
However, we measure the sample linear correlation between spot variance estimates and their square and find it to be approximately equal to $0.84$, thus signalling the presence of collinearity, which represents a violation of the OLS hypotheses. The problem of collinearity is typical of polynomial regressions and can be solved by transforming the regressors in equation (\ref{model2}) through the use of orthogonal polynomials, i.e., by performing an orthogonal polynomial regression (see \cite{orthoregr}). This way we are able to isolate the actual additional contribution of the square of variance estimates to the dynamics of the VIX index squared, if any. Accordingly, using the Gram-Schmidt algorithm, we transform the vector of spot variance estimates and the vector of their squared values into orthogonal vectors and estimate the following regression model:
\begin{equation}
\label{modelortho}
{\rm VIX}^2_t  =    q_{0,0}  + q_{0,1} \; Z^{(1)}_{t}  + q_{0,2} \; Z^{(2)}_{t },
\end{equation}
where $Z^{(1)}$ and $Z^{(2)}$ denote, respectively, the orthogonal transformations of the vector of the spot variance estimates and the vector of the squared spot variance estimates.
Clearly, the coefficients in equation (\ref{modelortho}) are not comparable to those those in (\ref{model2}). However, this is not relevant for our study, as we aim only at assessing the significance of the additional contribution of the squared variance estimates to the dynamics of the VIX squared, not at making inference of the coefficients in equation (\ref{model2}).

The results of the OLS estimation of the coefficients in (\ref{modelortho}) over the period 2006-2018 are reported in Table \ref{Table_2}. These results point out that the additional contribution of the squared spot variance is not statistically significant.
In order to interpret these results, we first need to recall that, from one side the class of polynomial models includes exponentially affine one as a subclass, from the other, in the presence of jumps, while the polynomial model gives rise to a quadratic correction (see Proposition \ref{prop-polynomial}), the exponentially affine model still ensures an affine link between the variance swap rate and the spot variance.
Therefore, as in Paragraph \ref{Affine_framework}, we have already ascertained that the exponentially affine framework is a suitable fit for S\&P500 data over the period 2006-2018, we deduce that the results in Table \ref{Table_2} confirm the adequacy of the exponentially affine framework in capturing the empirical features of S\&P500 data. In other words, Table \ref{Table_2} points towards the fact that the extension to the quadratic framework is not necessary to capture the empirical link between the variance swap rate and the spot variance.

%\begin{itemize}
%\item[-] in paragraph \ref{Affine_framework}, we have already ascertained that the exponentially affine framework is a suitable fit for S\&P500 data over the period 2006-2018;
%\item[-] in general, the class of polynomial models includes exponentially affine models as a subclass (see \cite{CKRT12});
%\item[-] in particular, in the presence of jumps, the exponentially affine model implies the existence of an affine link  between the variance swap rate and the spot variance, while the polynomial model implies the existence of a quadratic correction (see Proposition \ref{prop-polynomial}).
%\end{itemize}
Furthermore, Table \ref{Table_2} offers additional interesting insight. Recall that the computation of the VIX index is based on a daily sampling scheme. Thus, it is natural to ask whether the VIX index squared is adequately approximated by its asymptotic counterpart, namely the future expected quadratic variation. If this were not the case, one would observe a significant quadratic correction due to the discrete sampling, see Proposition \ref{sampling}. As this is not the case,
we infer that the continuous limit represents a very good approximation, thus providing empirical support to the numerical result by \cite{BJ08}.

\begin{table}[h!]
\begin{center}

\caption{Quadratic framework (\ref{modelortho}): estimation from S\&P500 data over the period 2006-2018 \\(p-values less than $10^{-4}$ are reported as zero).}
\label{Table_2}

\vspace{0.25cm}
\begin{tabular}{ c ||  c   c   c   c   c ||  c      }
framework & coeff. & estimate   & std. err. & t stat. & p value & $ R^2$  \\ \hline
quadratic & $q_{0,0}$ &0.0033 & 0.0001 & 27.1498 & 0  & 0.9585\\
       &  $q_{0,1}$ & 0.2464 &  0.0328  &  7.5065 & 0\\
                   &  $q_{0,2}$ & 0.0016 &  0.0471 &  0.0339 & 0.9730\\
\end{tabular}

\end{center}
\end{table}

As in the affine framework (see Paragraph \ref{Affine_framework}), we also analyze yearly sub-samples in order to evaluate if coefficient estimates are sensitive to events of distress. The ensuing estimation results are shown in Table \ref{Table_81}.
\begin{table}[h!]
\begin{center}

\caption{Quadratic framework (\ref{modelortho}): estimation from S\&P500 data over yearly subsamples \\ (p-values less than $10^{-4}$ are reported as zero).}
\label{Table_81}

\vspace{0.25cm}

\begin{tabular}{c ||     c  c  c   c   c ||  c      }
 \multicolumn{7}{ c }{ {Quadratic  framework}} \\
 \hline

  {year} & {coeff.} & {estimate}   & {std. err.} & {t stat.} & {p value}  & $ R^2$  \\ \hline

  2006	&	$q_{0,0} $	    &	    0.0017	&	    0.0007	&	  2.4286	 &	    0.0152	 &	0.4482	\\	
  	    &	$q_{0,1} $	    &	    0.6899	&	    0.3807	&	   2.4578   &	    0.0140	 &		    \\	
   	    &	$q_{0,2} $	    &	   -0.5633	&	    0.5811	&	  -0.9694  	 &	    0.3323	 &		    \\	\hline
 2007	&	$q_{0,0} $   	&	    0.0023	&	    0.0001	&	  19.2127   	&	    0	&	    0.6695	\\	
 	    &	$q_{0,1} $	    &	    0.2461	&	    0.0486	&	  5.0580    	&	    0	&		\\	
 	    &	$q_{0,2} $	    &	    -0.3650	&	    0.0543	&	    -6.7178  	&	    0	&		\\	\hline
 2008	&	$q_{0,0} $	    &	   0.0043	&	    0.0030	&	    1.4330   	&	    0.1518	&	0.9799    	\\	
 	    &	$q_{0,1} $	    &	    0.2366	&	    0.0290	&	    8.1595  	&	    0	&		\\	
 	    &	$ q_{0,2} $	    &	   0.0602	&	    0.1030	&	    0.5846  	&	    0.5588	&		\\	\hline
 2009   &	$ q_{0,0} $	    &	   0.0047	&	    0.0002	&	 20.4214  	&	    0	&	     0.8720	\\	
 	    &	$ q_{0,1} $	    &	    0.1540	&	    0.0510	&	  3.0196    &	    0.0025	&		\\	
 	    &	$q_{0,2} $	    &	   -0.3213	&	    0.0665	&	  -4.8310  	&	    0	&		\\	\hline
 2010	&	$ q_{0,0} $	    &	    0.0037	&	    0.0001	&	  30.3236  	&	    0	&	    0.8377	\\	
 	    &	$ q_{0,1} $	    &	    0.0257 	&	    0.0118	&	  2.1780   	&	    0.0294	&		\\	
 	    &	$ q_{0,2} $	    &	   -0.1861	&	    0.0560	&	  -3.3243  	&	    0.0009	&		\\	\hline	
 2011	&	$ q_{0,0} $	    &	    0.0037	&	    0.0030	&	   1.2330  &	    0.2175	&	    0.9298	\\	
 	    &	$ q_{0,1} $	    &	    0.0426	&	    0.0164	&	   2.5976  	&	    0.0094	&		\\	
 	    &	$q_{0,2} $	    &	   -0.3284	&	    0.1004	&	  -3.2707  	&	    0.0011	&		\\	\hline
 2012	&	$q_{0,0} $	    &	    0.0032	&	    0.0008	&	   3.7571     	&	    0	&	    0.3196	\\	
 	    &	$ q_{0,1} $	    &	    0.1534	&	    0.0337 &	  4.5479 	&	    0	&		\\	
 	    &	$q_{0,2} $	    &	    0.0099	&	    0.2435	&	  0.0407    	&	    0.9676	&		\\	\hline		
 2013	&	$ q_{0,0} $	    &	    0.0011	&	    0.0003	&	  3.7472 	&	    0.0002	&	   0.4961  	  	\\	
 	    &	$ q_{0,1} $	    &	    0.3715	&	    0.1235	&	  2.1682   	&	    0.0026	&		\\	
 	    &	$ q_{0,2} $	&	    -0.3151 &	    0.0872	&	  -3.6135	 	&	    0	&		\\	\hline	
 2014	&	$ q_{0,0} $	    &	    0.0053	&	    0.0014	&	    3.2852	&	    0.0002	&	    0.7387	\\	
 	    &	$ q_{0,1} $	    &	    1.2842	&	    0.5923	&	    2.0560	&	    0.0301	&		\\	
 	    &	$q_{0,2} $	&	   0.7698	&	    0.4071	&	   1.8909	&	    0.0586	&		\\	\hline
 2015	&	$q_{0,0} $	   &	    0.0025	&	    0.0001	&	    23.5592   	&	    0	&	    0.8896	\\	
 	    &	$q_{0,1} $	   &	    0.0283	&	    0.0035	&	  8.0547     	&	    0 	&		\\	
 	    &	$q_{0,2} $   &	   -0.0919	&	    0.0195	&	    -4.7018 	&	    0 	&		\\	\hline	
 2016	&	$q_{0,0} $	   &	    0.0023	&	    0.0001	&	    16.8683      	&	    0	&	    0.8668  	\\	
 	    &	$q_{0,1} $	   &	    0.1083	&	    0.0407	&	  2.6609  	&	         0.0078	&		\\	
 	    &	$q_{0,2} $   &	   -0.2189	&	    0.0865	&	  -2.5289   	&	    0.0114	&		\\	\hline	
 2017	&	$q_{0,0} $	   &	    0.0054	&	    0.0018	&	   3.0369     	&	    0.0024	&	    0.3295 	\\	
       	&	$q_{0,1} $	   &	    2.7480	&	    0.7214	&	  3.8095    	&	    0.0001	&		\\	
     	&	$q_{0,2} $   &	   -1.8918	&	    0.4822	&	  -3.9233   	&	    0.0001	&		\\	\hline	
 2018	&	$q_{0,0} $	   &	    0.0029	&	    0.0003	&	  10.0292   	&	    0    	&	    0.9140	\\	
 	    &	$q_{0,1} $	   &	    0.2417	&	    0.1191	&	  2.0294  	&	    0.0424	&		\\	
 	    & 	$q_{0,2} $   &	    0.0377	&	    0.1037	&	  0.3636   	&	    0.7162	&		\\	
\end{tabular}

\end{center}
 \end{table}
Note that the quadratic term is not statistically significant in 2006, 2008, 2012, 2014 and 2018. Based on Figure \ref{dataset}, panel b), and the detailed analysis in \cite{HX19}, these years appear truly different, in terms of the state of the financial market. For instance, during 2008 and 2018, the VIX exhibits
spikes, related, respectively, to the global financial crisis and  the `China-US trade war'. In contrast, 2006 and 2012 do not experience relevant economic events, see \cite{HX19}. The year 2014 represents an intermediate situation, where the VIX is almost flat until
the end of October, when a cluster of spikes  arises due to the end of quantitative easing policy by the Federal Reserve in the US. Thus, it is hard to attribute the statistical significance of the quadratic terms in Table \ref{Table_81} to events of financial distress. 

Focusing on the frequency of price jumps in Figure  \ref{boh}, panel b), we highlight that 2006, 2014, and 2018 show a relatively low
percentage of days with jumps. Thus, keeping in mind the result in Proposition \ref{prop-polynomial}, the non-significance of the quadratic coefficient in these years could be linked to the low percentage of days with jumps. The number of jumps in 2012 seems not coherent with this interpretation of the results in Table \ref{Table_81}. Indeed, the quadratic term is not statistically significant in 2012, despite the fact that the percentage of days with jumps in 2012 is the second largest after 2011. However, 2012 can be deemed as an atypical year, in terms of market liquidity. In 2012 a series of important expansionary monetary policies were started by central banks to respond to the Euro-zone debt crisis and its international ramifications. These include the decision by the European Central Bank to cut its rates in multiple steps and to start a long-term refinancing operation (LTRO) during the first trimester of 2012, and the decision by the US Federal Open Market Committee to start a quantitative easing
in September and to increase it in December of 2012. Thus, the year 2012 is characterized by an atypical number of positive jumps in response to this new paradigm of `Infinity Quantitative Easing', that massively increased the market liquidity. %Finally, note that the year 2008 shows a relative low number of jumps, despite the outbreak of the global financial crisis during that year. This is mainly because the global crisis started in September, so that its effects affected only the last quarter of 2008.

\subsubsection{Polynomial framework}
\label{Polynomial_framework}

In the last paragraph, the polynomial framework (\ref{feasiblepoly2}) is analyzed. Before fitting this model, we examine the sample correlation matrix of the regressors, which is shown in Table \ref{Table_3}.
\begin{table}[h!]

\caption{Sample correlation matrix of the regressors of the polynomial form (\ref{feasiblepoly2}) over the period 2006-2018. }
\label{Table_3}

\vspace{0.25cm}

\begin{center}
\begin{tabular}{ c | c   c   c   c   c      }
   & $ \widehat V_{t} $ & $ \widehat V^2_{t} $ & $ X_t $ & $ X^2_t $  & $ X_t\widehat V_{t} $  \\ \hline
$ \widehat V_{t} $    &     1    &     &      &     &    \\
$ \widehat V^2_{t} $  &     0.8394   &  1  &      &     &     \\
$ X_t $             &    -0.3702   & -0.1814  &   1   &      &     \\
$ X^2_t $           &    -0.3615   & -0.1765  &   0.9997   &  1    &     \\
$ X_t \widehat V_{t}$ &     0.9992   &  0.8341  &  -0.3550   & -0.3465   &  1    \\
\end{tabular}

\end{center}
 \end{table}
This table provides empirical evidence of the existence of an almost perfect linear dependence between the log-price and its square, and between the spot variance and the product of the log-price and the spot variance. Moreover, the analysis conducted in Section \ref{Quadratic_framework} has already shown that the additional contribution of the squared spot variance estimates to the dynamics of the squared VIX is not significant. Thus, it remains only to evaluate the additional contribution of the log-price. This polynomial framework could then be
associated with a {\sl fully affine} form in both the log-return and the spot variance.

In this regard, recall that it is a well-known stylized fact that asset price series are non-stationary.
Indeed, the Augmented Dickey Fuller test, performed at the 90\% confidence level, confirms that our log-price series has a unit root.
To cope with the non-stationarity of the log-price series, we estimate the coefficients in equation (\ref{model2}) after replacing log-prices
with their detrended values, i.e., their values minus their sample mean. The estimation results are summarized in Table \ref{Table_4}. \\

\begin{table}[h!]
\begin{center}

\caption{Polynomial (fully affine)  framework: estimation from S\&P500 data over the period 2006-2018 \\ (p-values less than $10^{-4}$ are reported as zero; $p_{1,0} $ indicates the coefficient of the detrended price).}
\label{Table_4}

\vspace{0.25cm }
\begin{tabular}{ c ||  c | c | c | c | c ||  c      }
{framework} & {coeff.} & {estimate}   & {std. err.} & {t stat.} & {p value}  & ${R}^2$  \\ \hline
polynomial & $ p_{0,0} $ &0.0013 & 0.0003 & 4.9275 & 0  & 0.9592\\
(fully affine)         &  $p_{0,1} $ & 0.0725 &  0.0106  &  6.8284 & 0\\
          &  $p_{1,0} $ & -0.0015 &  0.0009  & -1.7653 & 0.0775\\
\end{tabular}

\end{center}
 \end{table}
Based on Table \ref{Table_4}, the contribution of the log-price is not statistically significant at the $95\%$ confidence level,
but only at  $90\%$ level.
Overall, this result confirms that the affine framework is sufficient to adequately fit for our sample.
 
Finally, it is worth evaluating if the additional contribution of the price in explaining the dynamics of the VIX index squared is statistically significant on yearly subsamples, i.e., under different economic scenarios. The results of the year-by-year estimation are summarized in Table \ref{Table_8} and are in line with the whole-sample results.
\begin{table}[h!!]

\caption{Polynomial (fully affine) framework: estimation from S\&P500 data over yearly subsamples \\ (p-values less than $10^{-4}$ are reported as zero; $p_{1,0} $ indicates the coefficient of the detrended price).}\label{Table_8}

\vspace{0.25cm}
\begin{center}
\begin{tabular}{  c ||  c   c   c   c   c | | c      }

 \multicolumn{7}{ c }{ {Polynomial (fully affine) framework}} \\
 \hline

  {year} & {coeff.} & {estimate}   & {std. err.} & {t stat.} & {p value}  & ${R}^2$  \\
  \hline
2006	&	$p_{0,0} $	&	    0.0009	&	    0.0004	&	    2.2500	&	    0.0244	&	0.6012	\\	
	&	$p_{0,1} $	&	    0.0240	&	    0.0105	&	    2.2850	&	    0.0223	&		\\	
	&	$p_{1,0} $	&	   -0.0043	&	    0.0019	&	   -2.3374	&	    0.0194	&		\\	\hline
2007	&	$p_{0,0} $	&	    0.0016	&	    0.0004	&	    4.1944	&	    0.0000	&	    0.6699	\\	
	&	$p_{0,1} $	&	    0.0712	&	    0.0163	&	    4.3566	&	    0.0000	&		\\	
	&	$p_{1,0} $	&	    0.0054	&	    0.0031	&	    1.7258	&	    0.0844	&		\\	\hline
2008	&	$p_{0,0} $	&	   0.0019	&	    0.0011	&	   1.7273	&	    0.0841	&	    0.9796	\\	
	&	$p_{0,1} $	&	    0.0501	&	    0.0159	&	    3.1520	&	    0.0016	&		\\	
	&	$ p_{1,0} $	&	   -0.0322	&	    0.0084	&	   -3.8116	&	    0.0001	&		\\	\hline
2009    &	$ p_{0,0} $	&	   0.0038	&	    0.0013	&	   2.8957	&	    0.0038	&	    0.9093	\\	
	&	$ p_{0,1} $	&	    0.0728	&	    0.0188	&	    3.8755	&	    0.0001	&		\\	
	&	$p_{1,0} $	&	   -0.0164	&	    0.0041	&	   -3.9743	&	    0.0001	&		\\	\hline
2010	&	$ p_{0,0} $	&	   0.0020	&	    0.0007	&	   2.8571	&	    0.0043	&	    0.8291	\\	
	&	$ p_{0,1} $	&	    0.0428	&	    0.0149	&	    2.8654	&	    0.0042	&		\\	
	&	$ p_{1,0} $	&	   -0.0150	&	    0.0043	&	   -3.5235	&	    0.0004	&		\\	\hline
2011	&	$ p_{0,0} $	&	   0.0056	&	    0.0033	&	   1.6970	&	    0.0897	&	    0.9338	\\	
	&	$ p_{0,1} $	&	    0.0567	&	    0.0251	&	    2.2543	&	    0.0242	&		\\	
	&	$p_{1,0} $	&	   -0.0402	&	    0.0111	&	   -3.6063	&	    0.0003	&		\\	\hline
2012	&	$p_{0,0} $	&	0.0014	&	    0.0005	&	   2.8000	&	    0.0051	& 0.3944	   \\	
	&	$ p_{0,1} $	&	    0.0216	&	    0.0068	&	    3.1747	&	    0.0015	&		\\	
	&	$ p_{1,0} $	&	 0.0003	&	    0.0003	&	    0.8642	&	    0.3875	&	     	   	\\	\hline
2013	&	$ p_{0,0} $	&	    0.0013	&	    0.0001	&	   17.5755	&	         0	&	    0.4959	\\	
	&	$ p_{0,1} $	&	    0.0296	&	    0.0068	&	    4.3350	&	    0	&		\\	
	&	$ p_{1,0} $	&	    0.0003	&	    0.0004	&	    0.7629	&	    0.4455	&		\\	\hline
2014	&	$ p_{0,0} $	&	    0.0014	&	    0.0004	&	    3.2852	&	    0.0010	&	    0.5516	\\	
	&	$ p_{0,1} $	&	    0.0384	&	    0.0187	&	    2.0560	&	    0.0398	&		\\	
	&	$p_{1,0} $	&	   -0.0012	&	    0.0020	&	   -0.5993	&	    0.5490	&		\\	\hline
2015	&	$p_{0,0} $	&	    0.0101	&	    0.0016	&	    6.3290	&	    0 	&	    0.8942	\\	
	&	$p_{0,1} $	&	    0.0117	&	    0.0030	&	    3.8894	&	    0.0001	&		\\	
	&	$p_{1,0} $	&	   -0.0305	&	    0.0057	&	   -5.3606	&	    0 	&		\\	\hline
2016	&	$p_{0,0} $	&	    0.0040	&	    0.0008	&	    4.7158	&	    0 	&	    0.8996	\\	
	&	$p_{0,1} $	&	    0.0350	&	    0.0040	&	    8.7464	&	         0	&		\\	
	&	$p_{1,0} $	&	   -0.0088	&	    0.0027	&	   -3.2395	&	    0.0012	&		\\	\hline
2017	&	$p_{0,0} $	&	    0.0018	&	    0.0003	&	    6.2120	&	    0	&	    0.4695	\\	
	&	$p_{0,1} $	&	    0.0155	&	    0.0048	&	    3.2280	&	    0.0012	&		\\	
	&	$p_{1,0} $	&	   -0.0021	&	    0.0006	&	   -3.5198	&	    0.0004	&		\\	\hline
2018	&	$p_{0,0} $	&	    0.0054	&	    0.0021	&	    2.5861	&	    0.0097	&	    0.9133	\\	
	&	$p_{0,1} $	&	    0.0489	&	    0.0074	&	    6.6009	&	    0 &		\\	
	&	$p_{1,0} $	&	   -0.0075	&	    0.0036	&	   -2.1090	&	    0.0349	&		\\	
\end{tabular}

\end{center}

\end{table}
\begin{remark}
The study of Section \ref{Empirical} can be performed also using alternative spot variance estimators to reconstruct the series $\widehat V^m_{t_i}$. In fact, we have replicated the empirical study using the localized version of the two-scale realized variance and bipower variation (see Chapter 8 in \cite{aj}), in place, respectively, of the Fourier estimator by \cite{MM09} and the Fourier estimator by \cite{CT15}. The corresponding results are perfectly in line with those in Tables $1-7$, both in terms of the significance of the coefficient estimates and the $R^2$ values.
\end{remark}

\section{Conclusions}
\label{Conclusions}

This paper provides empirical evidence that S\&P500 data over the period 2006-2018 are coherent with the exponentially affine framework, introduced by \cite{KMKV2011}, who analytically prove the existence of an affine relationship between the expected future variance, i.e., the variance swap rate, and the spot variance. This paper collects empirical evidence that this affine relationship fits the data overwhelmingly well, with statistically significant coefficients and an $R^2$ larger than $0.95$.
Further, this paper provides empirical evidence that the daily sampling used to compute the actual variance swap rates
is frequent enough to erase the quadratic correction due to discrete sampling. The quadratic correction is expected within the polynomial framework, which includes the exponentially affine framework as a special case. This empirical non-parametric result confirms the result by  \cite{BJ08}, which was obtained on data simulated from four parametric models belonging to the exponentially affine class.

The paper focuses also on yearly subsamples, in order to evaluate the sensitivity to events of financial distress. Empirical results on yearly subsamples are more nuanced. In particular, it emerges that the exponentially affine framework could be rejected in 2008 and 2011. These two
years include the outbreaks of two global financial crisis sparked, respectively, by the American housing market and the sovereign debt in the Euro area. Models in the exponentially mean-reverting variance framework seem more adequate to capture the features of empirical data in those two years of financial distress. Finally, the significance of the quadratic coefficients in years with a relatively large number of price jumps supports the use of polynomial models in the presence of more frequent jumps.

%\section*{\textcolor{red}{Acknowledgement(s)}}
\section*{{Funding}}

S.Scotti acknowledges financial support from the Institut Europlace de Finance grant `Clusters and Information Flow: Modelling, Analysis and Implications' and from a visiting grant received by the University of Florence. \\
Part of the research was completed while M.E.Mancino was visiting the Université de Paris under a visiting funding scheme, which the author acknowledges. \\
%The authors acknowledge financial support from Scuola Normale superiore, 

\bibliographystyle{rQUF}
\bibliography{ref4}

\begin{thebibliography}{68}
\providecommand{\natexlab}[1]{#1}
\providecommand{\noopsort}[1]{}
\providecommand{\printfirst}[2]{#1}
\providecommand{\singleletter}[1]{#1}
\providecommand{\switchargs}[2]{#2#1}

\bibitem[\protect\citeauthoryear{Ackerer and Filipovic}{2020}]{AF16}
Ackerer, D. and Filipovic, D., Linear credit risk models. {\itshape Finance and
  Stochastics}, 2020, \textbf{24}, 169--214.

\bibitem[\protect\citeauthoryear{Ackerer {\itshape{et~al.}}}{2018}]{AFP18}
Ackerer, D., Filipovic, D. and Pulido, S., The Jacobi stochastic volatility
  model. {\itshape Finance and Stochastics}, 2018, \textbf{22}, 667--700.

\bibitem[\protect\citeauthoryear{A\"it-Sahalia and Jacod}{2014}]{aj}
A\"it-Sahalia, Y. and Jacod, J., High-frequency financial econometrics.
  {\itshape Princeton University Press}, 2014.

\bibitem[\protect\citeauthoryear{A\"it-Sahalia and Xiu}{2019}]{AX19}
A\"it-Sahalia, Y. and Xiu, D., A Hausman test for the presence of noise in high
  frequency data. {\itshape Journal of Econometrics}, 2019, \textbf{211},
  176--205.

\bibitem[\protect\citeauthoryear{Al\`os {\itshape{et~al.}}}{2007}]{ELV07}
Al\`os, E., Leon, J.A. and Vives, J., On the short-time behaviour of the
  implied volatility for jump-diffusion models with stochastic volatility.
  {\itshape Finance and Stochastics}, 2007, \textbf{11}, 571--589.

\bibitem[\protect\citeauthoryear{Andersen {\itshape{et~al.}}}{2001}]{ABDE01}
Andersen, T., Bollerslev, T., Diebold, F. and Ebens, H., The distribution of
  realized stock return volatility. {\itshape Journal of Financial Economics},
  2001, \textbf{61}, 43--76.

\bibitem[\protect\citeauthoryear{Bakshi {\itshape{et~al.}}}{1997}]{BCC97}
Bakshi, G., Cao, C. and Chen, Z., Empirical performance of alternative option
  pricing models. {\itshape The Journal of Finance}, 1997, \textbf{52},
  2003--2049.

\bibitem[\protect\citeauthoryear{Barndorff-Nielsen and Shephard}{2001}]{bnsmod}
Barndorff-Nielsen, O.E. and Shephard, N., Non‐Gaussian
  Ornstein–Uhlenbeck‐based models and some of their uses in financial
  economics. {\itshape Journal of the Royal Statistical Society}, 2001,
  \textbf{63}, 167--241.

\bibitem[\protect\citeauthoryear{Barndorff-Nielsen and
  Shephard}{2002{\natexlab{a}}}]{bns}
Barndorff-Nielsen, O.E. and Shephard, N., Estimating quadratic variation using
  realized variance. {\itshape Journal of Applied Econometrics},
  2002{\natexlab{a}}, \textbf{17}, 457--477.

\bibitem[\protect\citeauthoryear{Barndorff-Nielsen and
  Shephard}{2002{\natexlab{b}}}]{BarnShe2002}
Barndorff-Nielsen, O. and Shephard, N., Econometric analysis of realised
  volatility and its use in estimating stochastic volatility models. {\itshape
  Journal of the Royal Statistical Society. Series B (Statistical
  Methodology)}, 2002{\natexlab{b}}, \textbf{64}, 253–280.

\bibitem[\protect\citeauthoryear{Bates}{1996}]{Bates96}
Bates, D.S., Jumps and stochastic volatility: exchange rate processes implicit
  in Deutsche mark options. {\itshape The Review of Financial Studies}, 1996,
  \textbf{9}, 69--107.

\bibitem[\protect\citeauthoryear{Bayer
  {\itshape{et~al.}}}{2016}]{bayer2016pricing}
Bayer, C., Friz, P. and Gatheral, J., Pricing under rough volatility. {\itshape
  Quantitative Finance}, 2016, \textbf{16}, 887--904.

\bibitem[\protect\citeauthoryear{Beckers}{1980}]{beckers1980constant}
Beckers, S., The constant elasticity of variance model and its implications for
  option pricing. {\itshape The Journal of Finance}, 1980, \textbf{35},
  661--673.

\bibitem[\protect\citeauthoryear{Benth}{2011}]{Benth11}
Benth, F.E., The stochastic volatility model of Barndorff-Nielsen and Shephard
  in commodity markets. {\itshape Mathematical Finance}, 2011, \textbf{21},
  595--625.

\bibitem[\protect\citeauthoryear{Bernis {\itshape{et~al.}}}{2019}]{BSS19}
Bernis, G., Brignone, R., Scotti, S. and Sgarra, C., A Gamma Ornstein-Uhlenbeck
  Model Driven by a Hawkes Process. {\itshape SSRN paper n.3370304}, 2019.

\bibitem[\protect\citeauthoryear{Broadie and Jain}{2008}]{BJ08}
Broadie, M. and Jain, A., The effects of jumps and discrete sampling on
  volatility and variance swaps. {\itshape International Journal of Theoretical
  and Applied Finance}, 2008, \textbf{11}, 761-- 797.

\bibitem[\protect\citeauthoryear{Callegaro {\itshape{et~al.}}}{2017}]{CFP17}
Callegaro, G., Fiorin, L. and Pallavicini, A., Quantization goes polynomial.
  {\itshape ArXiv paper n.1710.11435}, 2017.

\bibitem[\protect\citeauthoryear{Carr and Lee}{2008}]{carr2008robust}
Carr, P. and Lee, R., Robust replication of volatility derivatives. {\itshape
  SSRN paper n.1108429}, 2008.

\bibitem[\protect\citeauthoryear{Carr and Sun}{2007}]{CS07}
Carr, P. and Sun, J., A new approach for option pricing under stochastic
  volatility. {\itshape Review of Derivatives Research}, 2007, \textbf{10},
  87--150.

\bibitem[\protect\citeauthoryear{Carr and Wu}{2006}]{CW06}
Carr, P. and Wu, L., A tale of two indices. {\itshape The Journal of
  Derivatives}, 2006, \textbf{13}, 13--29.

\bibitem[\protect\citeauthoryear{Carr and Wu}{2008}]{CW08}
Carr, P. and Wu, L., Variance risk premiums. {\itshape The Review of Financial
  Studies}, 2008, \textbf{22}, 1311--1341.

\bibitem[\protect\citeauthoryear{CBOE}{2019}]{VIXWhitePaper}
CBOE, VIX White paper. {\itshape Available at:
  www.cboe.com/micro/vix/vixwhite.pdf}, 2019.

\bibitem[\protect\citeauthoryear{Comte and Renault}{1998}]{ComteRenault}
Comte, F. and Renault, E., Long memory in continuous time stochastic volatility
  models. {\itshape Mathematical Finance}, 1998, \textbf{8}, 291--323.

\bibitem[\protect\citeauthoryear{Corsi {\itshape{et~al.}}}{2010}]{cpr}
Corsi, F., Pirino, D. and Renò, R., Threshold bipower variation and the impact
  of jumps on volatility forecasting. {\itshape Journal of Econometrics}, 2010,
  \textbf{159}, 276--288.

\bibitem[\protect\citeauthoryear{Cox {\itshape{et~al.}}}{1985}]{CIR85}
Cox, J., Ingersoll, J. and Ross, S., A theory of the term structure of interest
  rates. {\itshape Econometrica}, 1985, \textbf{53}, 385--407.

\bibitem[\protect\citeauthoryear{Cuchiero}{2011}]{Cuchiero-PhD}
Cuchiero, C., Affine and polynomial processes. {\itshape Ph.D. thesis (ETH
  Zurich)}, 2011.

\bibitem[\protect\citeauthoryear{Cuchiero}{2018}]{C18}
Cuchiero, C., Polynomial processes in stochastic portfolio theory. {\itshape
  Forthcoming in Stochastic processes and their applications}, 2018.

\bibitem[\protect\citeauthoryear{Cuchiero {\itshape{et~al.}}}{2012}]{CKRT12}
Cuchiero, C., Keller-Ressel, M. and Teichmann, J., Polynomial processes and
  their applications to mathematical finance. {\itshape Finance and
  Stochastics}, 2012, \textbf{16}, 711--740.

\bibitem[\protect\citeauthoryear{Cuchiero {\itshape{et~al.}}}{2018}]{CLSF18}
Cuchiero, C., Larsson, M. and Svaluto-Ferro, S., Polynomial jump-diffusions on
  the unit simplex. {\itshape Annals of Applied Probability}, 2018,
  \textbf{28}, 2451--2500.

\bibitem[\protect\citeauthoryear{Cuchiero and Teichmann}{2013}]{CT13}
Cuchiero, C. and Teichmann, J., Path properties and regularity of affine
  processes on general state spaces. {\itshape in S\'eminaire de Probabilit\'es
  XLV, Springer}, 2013.

\bibitem[\protect\citeauthoryear{Cuchiero and Teichmann}{2015}]{CT15}
Cuchiero, C. and Teichmann, J., Fourier transform methods for pathwise
  covariance estimation in the presence of jumps. {\itshape Stochastic
  Processes and Their Applications}, 2015, \textbf{125}, 116--160.

\bibitem[\protect\citeauthoryear{Dawson and Li}{2006}]{DL06}
Dawson, D. and Li, Z., Skew convolution semigroups and affine Markov processes.
  {\itshape The Annals of Probability}, 2006, \textbf{34}, 1103--1142.

\bibitem[\protect\citeauthoryear{Delbaen and Schachermayer}{1994}]{DS94}
Delbaen, F. and Schachermayer, W., A general version of the fundamental theorem
  of asset pricing. {\itshape Mathematische Annalen}, 1994, \textbf{300},
  463–520.

\bibitem[\protect\citeauthoryear{Dickey and Fuller}{1979}]{DFtest}
Dickey, D.A. and Fuller, W.A., Distribution of the estimators for
  autoregressive time series with a unit root. {\itshape Journal of the
  American Statistical Association}, 1979, \textbf{74}, 427–443.

\bibitem[\protect\citeauthoryear{Duffie {\itshape{et~al.}}}{2003}]{DufFilSch}
Duffie, D., Filipovic, D. and Schachermayer, W., Affine processes and
  applications in finance. {\itshape The Annals of Applied Probability}, 2003,
  \textbf{13}, 984--1053.

\bibitem[\protect\citeauthoryear{Duffie {\itshape{et~al.}}}{2000}]{DufPanSin}
Duffie, D., Pan, J. and Singleton, K., Transform analysis and asset pricing for
  affine jump diffusions. {\itshape Econometrica}, 2000, \textbf{68},
  1343--1376.

\bibitem[\protect\citeauthoryear{El~Euch {\itshape{et~al.}}}{2019}]{ElEuch}
El~Euch, O., Gatheral, J. and Rosenbaum, M., Roughening Heston. {\itshape
  Risk}, May 2019, pp. 84--89.

\bibitem[\protect\citeauthoryear{El~Euch and
  Rosenbaum}{2019}]{el2019characteristic}
El~Euch, O. and Rosenbaum, M., The characteristic function of rough Heston
  models. {\itshape Mathematical Finance}, 2019, \textbf{29}, 3--38.

\bibitem[\protect\citeauthoryear{Eraker}{2004}]{Eraker04}
Eraker, B., Do stock prices and volatility jump? Reconciling evidence from spot
  and option prices. {\itshape The Journal of Finance}, 2004, \textbf{59},
  1367–1403.

\bibitem[\protect\citeauthoryear{Filipovic}{2001}]{F01}
Filipovic, D., A general characterization of one factor affine term structure
  models. {\itshape Finance and Stochastics}, 2001, \textbf{5}, 389--412.

\bibitem[\protect\citeauthoryear{Filipovic {\itshape{et~al.}}}{2016}]{FGM16}
Filipovic, D., Gourier, E. and Mancini, L., Quadratic variance swap models.
  {\itshape Journal of Financial Economics}, 2016, \textbf{119}, 44–68.

\bibitem[\protect\citeauthoryear{Filipovic and Mayerhofer}{2009}]{FM09}
Filipovic, D. and Mayerhofer, E., Affine diffusion processes: theory and
  applications. {\itshape Advanced financial modelling}, 2009, \textbf{8},
  1--40.

\bibitem[\protect\citeauthoryear{Foster and Nelson}{1996}]{FosterNelson}
Foster, D.P. and Nelson, D.B., Continuous record asymptotics for rolling sample
  variance estimators. {\itshape Econometrica}, 1996, \textbf{64}, 139–174.

\bibitem[\protect\citeauthoryear{Gatheral
  {\itshape{et~al.}}}{2018}]{gatheral2018volatility}
Gatheral, J., Jaisson, T. and Rosenbaum, M., Volatility is rough. {\itshape
  Quantitative Finance}, 2018, \textbf{18}, 933--949.

\bibitem[\protect\citeauthoryear{Granger and Newbold}{1974}]{Granger}
Granger, C. and Newbold, P., Spurious regressions in econometrics. {\itshape
  Journal of Econometrics}, 1974, \textbf{2}, 111--120.

\bibitem[\protect\citeauthoryear{Grasselli}{2016}]{G2016}
Grasselli, M., The 4/2 stochastic volatility model: a unified approach for the
  Heston and the 3/2 model. {\itshape Mathematical Finance}, 2016, \textbf{27},
  1013--1034.

\bibitem[\protect\citeauthoryear{Hagan
  {\itshape{et~al.}}}{2002}]{hagan2002managing}
Hagan, P., Kumar, D., Lesniewski, A. and Woodward, D., Managing smile risk.
  {\itshape The Best of Wilmott}, 2002, \textbf{1}, 249--296.

\bibitem[\protect\citeauthoryear{Heston}{1993}]{Heston}
Heston, S., A closed-form solution for options with stochastic volatility with
  applications to bond and currency options. {\itshape The Review of Financial
  Studies}, 1993, \textbf{6}, 327--343.

\bibitem[\protect\citeauthoryear{Horst and Xu}{2019}]{HX19}
Horst, U. and Xu, W., The microstructure of stochastic volatility models with
  self exciting jump dynamics. {\itshape ArXiv paper n.1911.12969}, 2019.

\bibitem[\protect\citeauthoryear{Hubalek {\itshape{et~al.}}}{2017}]{HKRS17}
Hubalek, F., Keller-Ressel, M. and Sgarra, C., Geometric Asian option pricing
  in general affine stochastic volatility models with jumps. {\itshape
  Quantitative Finance}, 2017, \textbf{17}, 873--888.

\bibitem[\protect\citeauthoryear{Hull and White}{1987}]{hull1987pricing}
Hull, J. and White, A., The pricing of options on assets with stochastic
  volatilities. {\itshape The Journal of Finance}, 1987, \textbf{42}, 281--300.

\bibitem[\protect\citeauthoryear{Jaber
  {\itshape{et~al.}}}{2019}]{jaber2019weak}
Jaber, E.A., Cuchiero, C., Larsson, M. and Pulido, S., A weak solution theory
  for stochastic Volterra equations of convolution type. {\itshape arXiv
  preprint arXiv:1909.01166}, 2019.

\bibitem[\protect\citeauthoryear{Jacquier
  {\itshape{et~al.}}}{2018}]{jacquier2018vix}
Jacquier, A., Martini, C. and Muguruza, A., On VIX futures in the rough Bergomi
  model. {\itshape Quantitative Finance}, 2018, \textbf{18}, 45--61.

\bibitem[\protect\citeauthoryear{Jarrow {\itshape{et~al.}}}{2013}]{Protter}
Jarrow, R., Kchia, Y., Larsson, M. and Protter, P., Discretely sampled variance
  and volatility swaps versus their continuous approximations. {\itshape
  Finance and Stochastics}, 2013, \textbf{17}, 305--324.

\bibitem[\protect\citeauthoryear{Jiao {\itshape{et~al.}}}{2017}]{JMS17}
Jiao, Y., Ma, C. and Scotti, S., Alpha-CIR model with branching processes in
  sovereign interest rate modeling. {\itshape Finance and Stochastics}, 2017,
  \textbf{21}, 789--813.

\bibitem[\protect\citeauthoryear{Jiao {\itshape{et~al.}}}{2019}]{JMSZ19}
Jiao, Y., Ma, C., Scotti, S. and Zhou, C., The Alpha-Heston stochastic
  volatility model. {\itshape https://arxiv.org/abs/1812.01914}, 2019.

\bibitem[\protect\citeauthoryear{Kallsen {\itshape{et~al.}}}{2011}]{KMKV2011}
Kallsen, J., Muhle-Karbe, J. and Voss, M., Pricing option on variance in affine
  stochastic volatility models. {\itshape Mathematical Finance}, 2011,
  \textbf{21}, 627--641.

\bibitem[\protect\citeauthoryear{Keller-Ressel}{2011}]{KR2011}
Keller-Ressel, M., Moment explosions and long-term behavior of affine
  stochastic volatility models. {\itshape Mathematical Finance}, 2011,
  \textbf{21}, 73--98.

\bibitem[\protect\citeauthoryear{Keller-Ressel
  {\itshape{et~al.}}}{2011}]{KRST2011}
Keller-Ressel, M., Schachermayer, W. and Teichmann, J., Affine processes are
  regular. {\itshape Probability Theory and Related Fields}, 2011,
  \textbf{151}, 591--611.

\bibitem[\protect\citeauthoryear{Li}{2011}]{Li2011}
Li, Z., Measure-Valued Branching Markov Processes. {\itshape Springer}, 2011.

\bibitem[\protect\citeauthoryear{Li and Ma}{2008}]{LM08}
Li, Z. and Ma, C., Catalytic discrete state branching models and related limit
  theorems. {\itshape Journal of Theoretical Probability}, 2008, \textbf{21},
  936--965.

\bibitem[\protect\citeauthoryear{Malliavin and Mancino}{2009}]{MM09}
Malliavin, P. and Mancino, M., A Fourier transform method for nonparametric
  estimation of multivariate volatility. {\itshape The Annals of Statistics},
  2009, \textbf{37}, 1983--2010.

\bibitem[\protect\citeauthoryear{Mancino and Recchioni}{2015}]{MaRec}
Mancino, M. and Recchioni, M., Fourier spot volatility estimator: asymptotic
  normality and efficiency with liquid and illiquid high-frequency data.
  {\itshape PLoS ONE}, 2015, \textbf{10}.

\bibitem[\protect\citeauthoryear{Mancino and Sanfelici}{2008}]{MaSan}
Mancino, M. and Sanfelici, S., Robustness of Fourier estimator of integrated
  volatility in the presence of microstructure noise. {\itshape Computational
  Statistics and Data Analysis}, 2008, \textbf{52}, 2966–2989.

\bibitem[\protect\citeauthoryear{Mykland and Zhang}{2006}]{MyklandZhang2006}
Mykland, P. and Zhang, L., Anova for diffusions. {\itshape Annals of
  Statistics}, 2006, \textbf{34}, 1931–1963.

\bibitem[\protect\citeauthoryear{Narula}{1979}]{orthoregr}
Narula, S., Orthogonal Polynomial Regression. {\itshape International
  Statistical Review}, 1979, \textbf{47}, 31--36.

\bibitem[\protect\citeauthoryear{Newey and West}{1987}]{NeweyWest}
Newey, W. and West, K., A simple, positive semi-definite, heteroskedasticity
  and autocorrelation consistent covariance matrix. {\itshape Econometrica},
  1987, \textbf{55}, 703--708.

\bibitem[\protect\citeauthoryear{Platen}{1997}]{32}
Platen, E., A non-linear stochastic volatility model. {\itshape Financial
  Mathematics Research Report No.FMRR005-97, Center for Financial Mathematics,
  Australian National University}, 1997.

\end{thebibliography}

\appendices
\section{Proofs}

\begin{proof}{(of Proposition \ref{prop-coeffcient-affine})}

Here we adopt, where no ambiguity arises, the parameter notation introduced in Lemma 4.2 of
\cite{KMKV2011} for the exponentially affine model.
Using the equation for $\Psi^V_u$ and $\Phi^V_u$ in Proposition \ref{KAL} and
%satisfies
%\begin{eqnarray*}
%\frac{\upartial \Psi^V_u}{\upartial t}(t) &=& \frac{1}{2}\gamma_1^{11} \left(\Psi^V_u(t)\right)^2 + \beta_1^1 \Psi^V_u(t) + \gamma_1^{22} u
% +\int \left( e^{x_1\Psi^V_u(t) +ux_2^2} -1 - \Psi^V_u(t) h(x_1)  \right) \kappa_1(dx) \\
%\Psi^V_u(0) &=& 0 \\
%\Phi^V_u(t) &=& \int_0^t  \left[\beta_0^1 \Psi^V_u(s) +\gamma_0^{22} u + \int\left( e^{x_1\Psi^V_u(s) +ux_2^2} -1 - \Psi^V_u(s) h(x_1)
%\right) \kappa_0(dx) \right] ds
%\end{eqnarray*}
differentiating the two equations with respect to $u$, we have
\begin{eqnarray*}
\frac{\upartial^2 \Psi^V_u}{\upartial u \upartial t}(t) &=& \gamma_1^{11} \Psi^V_u(t) \frac{\upartial \Psi^V_u(t)}{\upartial u} + \beta_1^1
\frac{\upartial \Psi^V_u(t)}{\upartial u} + \gamma_1^{22}   \\
&& +\int_{\mathbb{R}^+\times \mathbb{R}} \left(   e^{x_1\Psi^V_u(t) +ux_2^2} \left( x_1 \frac{\upartial \Psi^V_u(t)}{\upartial u}
  + x_2^2 \right)   -  \frac{\upartial \Psi^V_u(t)}{\upartial u} h(x_1)  \right) \kappa_1({\rm d}x), \\
\frac{\upartial \Phi^V_u(t)}{\upartial u} &=& \int_0^t  \left[\beta_0^1 \frac{\upartial \Psi^V_u(s)}{\upartial u} +\gamma_0^{22} +
\int_{\mathbb{R}^+\times \mathbb{R}}
 \left(  e^{x_1\Psi^V_u(s) +ux_2^2} \left( x_1 \frac{\upartial \Psi^V_u(s)}{\upartial u}
  + x_2^2 \right)   -  \frac{\upartial \Psi^V_u(s)}{\upartial u} h(x_1)
\right) \kappa_0({\rm d}x) \right] {\rm d}s \,.
\end{eqnarray*}
Taking $u=0$ and recalling that $\Psi^V_0(t) = 0$ for all $t$, we obtain the relations satisfied by $(\Psi,\Phi)$, that read
\begin{eqnarray*}
\frac{\upartial \Psi}{\upartial t}(t) &=&  \beta_1^1 \Psi(t) + \gamma_1^{22}
+ \int_{\mathbb{R}^+\times \mathbb{R}}  \left(    x_1 \Psi(t)   + x_2^2
     - \Psi(t)  h(x_1)  \right) \kappa_1({\rm d}x), \\
\Phi(t) &=& \int_0^t  \left[\beta_0^1  \Psi(s) +\gamma_0^{22} + \int_{\mathbb{R}^+\times \mathbb{R}} \left(  x_1 \Psi(s)
  + x_2^2  -  \Psi(s) h(x_1)  \right) \kappa_0({\rm d}x) \right] {\rm d}s \,.
\end{eqnarray*}
Note that in our case $\gamma_0^{22}=0$, see (\ref{eq-general-model}). Then, splitting the integrals and recalling that $h(x)$ is a
truncating function, there exists non-negative parameters $(\widetilde\beta_1^1, \widetilde\beta_0^1)$, that is the parameters associated with
the truncating function $h(x)=x$,  such that:
\begin{eqnarray*}
\frac{\upartial \Psi}{\upartial t}(t) &=&  \widetilde\beta_1^1 \Psi(t) + \gamma_1^{22}
+ \int_{\mathbb{R}^+\times \mathbb{R}}  x_2^2 \;  \kappa_1({\rm d}x), \\
\Phi(t) &=& \int_0^t  \left[\widetilde\beta_0^1  \Psi(s) + \int_{\mathbb{R}^+\times \mathbb{R}}
   x_2^2 \; \kappa_0({\rm d}x) \right] {\rm d}s \,.
\end{eqnarray*}
Note that $\Psi$ solves a non-homogeneous linear differential equation with non-negative external  term
$ \gamma_1^{22}  + \int_{\mathbb{R}^+\times \mathbb{R}} x_2^2 \;  \kappa_1({\rm d}x)$.
This term is zero if and only if $\gamma_1^{22}=0$ and $ \kappa_1({\rm d}x)=0$, that is in the case of
the exponential Levy model. We deduce that $\Psi(s)>0$ for all positive $s$,
except for the exponential Levy model, for which volatility is constant and thus the stationary distribution is degenerate.
We now turn to $\Phi$ and assume $\Psi(s)>0$. Using the integral representation of $\Phi$, we easily obtain that $\Phi>0$ if and only if
$\widetilde\beta_0^1\neq 0$ or   $\kappa_0({\rm d}x) \neq 0$, see also  \cite{F01}.
%{\bf non mi torna, se $\widetilde\beta_0^1< 0$ potrebbe andare negativo: NO citare Filipovic 01, per $k_0\geq 0$}.
This is equivalent to assuming that the process $V$ is a continuous-state branching
process with immigration. Instead, in the case $\widetilde\beta_0^1= 0$ and   $\kappa_0({\rm d}x) = 0$, the process $V$ is a continuous-state branching
process without immigration. Without immigration continuous-state branching processes do not have a stationary distribution, see Theorem 3.20 and Corollary 3.21 in \cite{Li2011}.
 %That is in contrast with the hypothesis of the existence of a stationary distribution for the variance process.
\end{proof} 

\begin{proof}{(of Proposition \ref{exponential_mean_reverting_Prop})}

The quadratic variation of $X$ is rewritten as $[X]_t = [X^c]_t +\sum_{s\leq t} (\Delta X_s)^2$, where $X^c$ denotes the continuous part of
the log-price $X$. According to (\ref{eq-general-mean-reverting}), we have $[X^c]_t = \int_0^t V_s {\rm d}s$.
It is easy to show that the variance process $V$ is integrable using Gronwall's lemma, since the drift of the variance process
is affine and $Z$ is integrable by hypothesis.
We now focus on the jump contribution, $\sum_{s\leq t} (\Delta X_s)^2 = \sum_{s\leq t} (\Delta J_s)^2$. Recalling that the process $J$
is square-integrable, we obtain that the optional version of the quadratic variation $[X]_t$ is finite almost surely.
Introducing the predictable version $\langle X\rangle _t$ of the quadratic variation and recalling that the optional and the predictable version of
the quadratic variation differ by a martingale, we obtain that
$$
{  E}\left[ [X]_\tau \right] = {  E}\left[ \langle X\rangle_\tau   \right] =
{  E}\left[ \int_0^\tau V_s {\rm d}s + \int \int_0^\tau \zeta^2 \nu(d\zeta, {\rm d}s) \right] \,.
$$
Considering first the jump term, 
exploiting that $\nu$ is affine in the variance process $V$, we deduce that the jump part is affine in the expectation
of the integral of the variance process. Focusing now on the term ${  E}\left[ \int_0^\tau V_s {\rm d}s\right]$, we consider the integral version of the
SDE (\ref{eq-general-mean-reverting}), i.e.
$$
V_t -V_0 = \kappa \int_0^t (\theta-V_s) {\rm d}s + Z_t.
$$
Taking the expectation, we obtain that ${  E}\left[ \int_0^\tau V_s {\rm d}s \right] =  \kappa^{-1}
\left(\tau\kappa \theta +V_0 - {  E}\left[ V_\tau \right]\right) $ and ${  E}\left[ V_t \right]$ satisfies a linear ODE.
This result, combined with the previous result, proves that the expectation of the quadratic variation is an affine function of the initial spot
variance.
\end{proof} 

\begin{proof}{(of Proposition \ref{prop-polynomial})}

According to the characterization in Proposition 2.12 of \cite{CKRT12}, if $(X,V)$ is a $2$-polynomial process then
$$[X,X]_t= \int_0^t V_s {\rm d}s + \int_0^t \int \zeta^2 \nu({\rm d}\zeta, {\rm d} s) =: \int_0^t a(X_s, V_s) {\rm d}s,$$
where $a\in {\cal P}_2$. Then, the result for $E^{(x,v)}[[X,X]_t]$ follows from Theorem 3.2 in \cite{Cuchiero-PhD} and the application of the stochastic Fubini theorem.
\par\noindent
In particular, if we consider the quadratic variation of $X^c$, together with the evolution \eqref{eq-general-model}, we have that $[X^c]_t = \int_0^t V_s {\rm d}s$.
Taking the expectation and applying the stochastic Fubini theorem, we obtain ${  E}\left[[X^c]_t \right]= \int_0^t {  E}\left[V_s\right] {\rm d}s$.
Now, using the hypothesis that $(X,V)$ is $2$-polynomial, we see that the function $f(x,v)=v \in {\cal P}_1$, and integrating we obtain the result.
\end{proof} 

\begin{proof}{(of Proposition \ref{sampling})}

Using the definition of realized variance $RV^m_{[0,\tau]}$ given in (\ref{RV}),
%, and for simplicity consider a regular sampling for $[0,\tau]$.
%let $\pi_m$ such that for all $k=1,\ldots,m$ $t_k-t_{k-1}$ is one open day.
%We then consider the sequence
%$RV_{m}:= \sum_{k=1}^{m} \left( X_{t_k} - X_{t_{k-1}}\right)^2\,$,
%where $N(n)$ denotes the cardinality of $\pi^{(n)}$ and $\{t^{(n)}_k\}_k$ is the ordered sequence of times in $\pi^{(n)}$.
%By triangular inequality, we have that  $RV^{(n)}$ is a non decreasing sequence. Moreover it is bounded by the quadratic variation of $X$
%that is bounded by the semi-martingale hypothesis. Assuming that the step of partition $\vert\pi^{(n)}\vert$ decreases to zero, we have
%that $RV^{(n)}$ converges to the quadratic variation $[X]$.
we take the expected value of $RV^m$. Due to the finiteness of the sum, the expected value of $RV^m_{[0,\tau]}$ gives
$${  E}^{(x,v)}\left[RV^m_{[0,\tau]}\right]= \sum_{k=1}^{m} {  E}^{(x,v)} \left[ \left( X_{t_k} - X_{t_{k-1}}\right)^2 \right].$$
Noting that the function inside the expectation belongs to ${\cal P}_2$, we have that ${  E}\left[RV^m_{[0,\tau]}\right] $ belongs to ${\cal P}_2$.
\end{proof}

\end{document}